# Characterization of the Tail of the Distribution of Earthquake Magnitudes by combining the GEV and GPD descriptions of Extreme Value Theory

## V.F. Pisarenko[1], A. Sornette[2], D. Sornette[3] and M.V. Rodkin[4]

[1]International Institute of Earthquake Prediction Theory
and Mathematical Geophysics
Russian Ac. Sci., Profsoyuznaya 84/32, Moscow 117997, Russia

[2]ETH Zurich, Swiss Seismological Service
HPP P6.1, Hönggerberg, CH-8093 Zürich, Switzerland

[3]ETH Zurich, D-MTEC, Kreuzplatz 5
CH-8032 Zurich, Switzerland
University of California, Los Angeles, California 90095

[4]Geophysical Center of
Russian Ac. Sci. Molodezhnaya 3, Moscow 119206, Russia

**Abstract:** The present work is a continuation and improvement of the method suggested in [Pisarenko et al. 2008] for the statistical estimation of the tail of the distribution of earthquake sizes. The chief innovation is to combine the two main limit theorems of Extreme Value Theory (*EVT*) that allow us to derive the distribution of *T*-maxima (maximum magnitude occurring in sequential time intervals of duration *T*) for arbitrary *T*. This distribution enables one to derive any desired statistical characteristic of the future *T*-maximum. We propose a method for the estimation of the unknown parameters involved in the two limit theorems corresponding to the Generalized Extreme Value distribution (*GEV*) and to the Generalized Pareto Distribution (*GPD*). We establish the direct relations between the parameters of these distributions, which permit to evaluate the distribution of the *T*-maxima for arbitrary *T*. The duality between the *GEV* and *GPD* provides a new way to check the consistency of the estimation of the tail characteristics of the distribution of earthquake magnitudes for earthquake occurring over arbitrary time interval. We develop several procedures and check points to decrease the scatter of the estimates and to verify their consistency. We test our full procedure on the global Harvard catalog (1977-2006) and on the Fennoscandia catalog (1900-2005). For the global catalog, we obtain the following estimates: $\hat{M}_{max}$ = 9.53 ± 0.52; $\hat{Q}_{10}(0.97)$ =9.21 ± 0.20. For Fennoscandia, we obtain $\hat{M}_{max}$ = 5.76 ± 0.165; $\hat{Q}_{10}(0.97)$ =5.44 ± 0.073. The estimates of all related parameters for the *GEV* and *GPD*, including the most important form parameter, are also provided. We demonstrate again the absence of robustness of the generally accepted parameter characterizing the tail of the magnitude-frequency law, the maximum possible magnitude $M_{max}$, and study the more stable parameter $Q_T(q)$, defined as the *q*-quantile of the distribution of *T*-maxima on future interval of duration *T*.

### *1-Introduction*

The magnitude-frequency Gutenberg-Richter (*G-R*) law is the most documented and robust statistical law of seismology. It has been subjected to numerous investigations (see, e.g. Bird and Kagan, 2004; Cosentino et al., 1977; Kagan, 1991; 1996; 1999; 2002a; 2002b; Kijko and Sellevol 1989, 1992; Knopoff et al., 1982; Main et al., 1999; Ogata and Katsura, 1993; Pisarenko and Sornette, 2003; 2004; Sornette et al., 1996; Utsu, 1999; Wu, 2000; Pisarenko and Rodkin, 2007). For small and moderate magnitudes, and for large space-time volumes, the Gutenberg-Richter law is valid with a large degree of accuracy.

However, for the largest magnitudes, some more or less significant deviations of the distribution of earthquake magnitudes from the *G-R* law have been documented (see e.g. Pisarenko and Sornette (2004) and references therein). The intrinsic difficulty of the investigation of the largest magnitudes is the insufficient number of large earthquakes. Inevitably, the numerous proposals for novel models of the deviations of the tail of the distribution of earthquake magnitudes from the *G-R* suffer from large statistical uncertainty. As a consequence, the problem of finding an adequate description of the tail of the magnitude distribution cannot be considered as definitely settled. One of the best-known modifications of the *G-R* (Kagan, 1997; Kagan and Schoenberg, 2001; Bird and Kagan 2004) consists in multiplying the power law distribution of seismic moments (which corresponds to the *G-R* exponential distribution of magnitudes) by an exponential factor (also referred to as an exponential taper), which leads to a Gamma-distribution for seismic moments. The characteristic index in the exponential taper is often referred to as the "corner" moment, as it constitutes the typical magnitude beyond which the distribution departs significantly downward from the pure *G-R* law.

Rather than introducing a "soft" truncation of the *G-R* law, a different class of models assume that the *G-R* law holds up to a maximum magnitude $M_{max}$, beyond which no earthquake are observed (Cosentino et al., 1977; Dargahi-Noubary, 1983; Main et al., 1999; Pisarenko, 1991; Pisarenko et al., 1996)

$$F(x) = \begin{cases} 0; & x < m ; \\ [\,10^{-\beta m} - 10^{-\beta x}\,]\,/\,[\,10^{-\beta m} - 10^{-\beta M max}\,]; & m \le x \le M_{max} ; \\ 1; & x > M_{max} . \end{cases}$$

The parameter $M_{max}$ represents the *maximum possible earthquake size* in the region under study. This parameter plays a very important role in seismic risk assessment and in seismic hazard mitigation (see e.g. Bender and Perkins, 1993; Pisarenko et al., 1996; Kijko and Graham, 1998; Kijko et al., 2001). It should be noted that the truncated *G-R* ensures the finiteness of the mean seismic energy, whereas the *G-R* in its unlimited form corresponds to a regime with infinite seismic energy, which is, of course, an undesirable property of the model. The parameter $M_{max}$ is convenient for engineers and insurers: having a reliable estimate of $M_{max}$, it is comparatively easy to take adequate decisions on the construction standards of buildings or on insurance policy. As a consequence, the truncated *G-R* has undergone a wide dissemination. Unfortunately, all attempts so far for a reliable statistical estimation of $M_{max}$ in various seismic regions did not give satisfactory results due to large statistical scatter and lack of reliability of its estimates. Attempts to attribute a maximum magnitude to individual faults rather than to regions suffer from the same problems and in addition face the fact that many large earthquakes involve several faults or fault segments which are difficult if not impossible to determine in advance (Black et al., 2004; Ward, 1997).

In order to avoid these undesirable properties of the parameter $M_{max}$ we have suggested in [Pisarenko et al.2008] to use the more stable parameter $Q_T(q)$ – the *q*-quantile of the distribution of *T*-maximum in intervals of duration *T*. In the present paper, we refine this approach and establish important relations between the two main limit theorems of Extreme Value Theory (*EVT*), allowing us to characterize the distribution of future *T*-maximum for any desired *T*.

The paper is structured as follows. Section 2 presents the two main theorems of *EVT* and 3 Corollaries that we shall use in our estimation approach (the Corollaries are proved in the Appendix). Then, two approaches for the estimation of the parameters corresponding to the *GPD* and *GEV* distributions are defined and tested on synthetic catalogs. We explain in details the bootstrap methods and the statistical methods developed to (1) decrease the scatter of the estimated parameters and (2) quantify the remaining level of uncertainty. Section 3 presents the application of our method to the global worldwide Harvard catalog. Section 4 applies our method to the local catalog of Fennoscandia. Section 5 concludes.

## 2. The method of combined use of the two main theorems of Extreme Value Theory.

### 2.1 Statement of the two main theorems of extreme value theory

Extreme value theory (*EVT*) studies the limiting distribution of the maxima of iid rv (identically independently distributed random variables) as the sample size $n$ tends to infinity. Two main limit theorems constitute the statistical basis for applications of *EVT*: (i) the main limit theorem of *EVT* (proved by Frechet (1927) for Pareto-type limit distribution and by Fisher and Tippet (1928) for Weibull and Gumbel limit distributions) leading to the Generalized Extreme Value distribution (*GEV*) and (ii) the theorem by Gnedenko-Pickands-Balkema-Haan leading to the Generalized Pareto Distribution (*GPD*). We now state these two theorems in turn.

#### 2.1.1 First theorem

**Frechet(1927)-Fisher-Tippett(1928) theorem (FFT),** (see e.g. [Embrechts et al. 1997])

*Let $x_1, \ldots x_n, \ldots$ be iid rv with a continuous distribution function (DF) F(x);*
*Let $M_n = max(x_1, \ldots x_n)$ be the maximum of a sample of n such rv.*
*If there exist two series of numbers $a_n$, $b_n$, such that $(M_n - a_n)/b_n$ weakly converges to a **non-degenerate** rv with DF $\Phi(x)$, then up to a shift and to a scale change, $\Phi(x)$ can only take one of the three forms:*

*$\Phi_1(x|\xi) = exp(-1/x^{1/\xi})$, $x > 0$; $\xi > 0$; (Pareto type);*
*$\Phi_2(x|\xi) = exp(-|x|^{-1/\xi})$, $x < 0$; $\xi < 0$; $\Phi_2(x)=1$, $x \geq 0$; (Weibull type);*
*$\Phi_3(x) = exp(-exp(-x))$, $|x| < \infty$; (Gumbel type).*

These three distributions can be written in a unified form, using the notations $\mu$ and $\sigma$ to refer to the centering and scale parameters:

*$\Phi(x|\mu,\sigma,\xi) = exp\{ -[1+\xi(x-\mu)/\sigma]^{-1/\xi} \}$, $1+\xi(x-\mu)/\sigma > 0$; $\sigma > 0$; $0 <|\xi| < \infty$; $|\mu| < \infty$. (1)*
*$\Phi(x|\mu,\sigma,0 ) = exp\{ -exp[-(x-\mu)/\sigma] \};$ (corresponding to $\xi = 0$)*

The form parameter $\xi$ varies from minus infinity to plus infinity. Its sign determines the domain of definition of the rv:

*$\xi > 0$; $x \geq \mu - \sigma/\xi$; Pareto;*
*$\xi < 0$; $x \leq \mu - \sigma/\xi$; Weibull;*
*$\xi = 0$; $|x| < \infty$; Gumbel.*

The distribution (1) is called the *Generalized Extreme Value* (*GEV*) distribution.

### 2.1.2 Second theorem

In order to formulate the second theorem, we introduce the right limit point $x_F$ of the distribution $F(x)$ (which is infinity for unbounded rv) and the excess function $F_H(x)$ (defined as the conditional distribution over the threshold $H$):

$$x_F = sup\{ x\colon F(x) < 1 \}, \qquad (2)$$
$$F_H(x) = P\{ X - H < x \mid X > H \}, \ x \geq 0. \qquad (3)$$

**Gnedenko-Pickands-Balkema-Haan theorem (GPBH),** (see e.g. [Bassi et al. 1998])

*Let $x_1, x_2, ...x_n$ be iid rv with the continuous distribution function $F(x)$ and excess function $F_H(x)$. We denote $M_n = max(x_1, x_2, ...x_n)$ the maximum of a sample of $n$ such rv.. Suppose that there exist some normalizing constants $c_n$, $d_n$, such that the normalized maximum $(M_n - d_n)/c_n$ weakly converges to a non-degenerate rv. Then, there exists a non-negative function $s(H)$ such that*

$$\lim_{H \uparrow x_F} \quad \sup_{0 \leq x \leq x_F - H} | F_H(x) - G(x \mid \gamma, s(H)) | = 0, \qquad (4)$$

*where $G(x \mid \gamma, s)$ is the Generalized Pareto Distribution (GPD):*

$$G(x \mid \gamma, s) = 1 - (1 + \gamma x/s)^{-1/\gamma}, \quad 0 < |\gamma| < +\infty ; \ s > 0, \qquad (5)$$
$$G(x \mid 0, s) = 1 - exp(-x/s); \qquad (corresponding\ to\ \gamma = 0).$$

*The domain of definition of G is determined by the sign of $\gamma$ :*

$$x \geq 0 \quad for\ \gamma \geq 0 ; \qquad 0 \leq x \leq - s / \gamma \quad for\ \gamma < 0. \qquad (6)$$

This second theorem implies that the limit of any excess function $F_H(x)$, which satisfies the conditions of this theorem, is an universal distribution – the Generalized Pareto Distribution (GPD).

### 2.2 Duality between the two main theorems of extreme value theory

These two limit theorems show that information on the distribution of extremes can be gathered in two ways: (1) by measuring the maximum of a sample whose size $n$ goes to infinity, or (2) by recording the excess function $F_H(x)$ of that sample when increasing the threshold $H$ to its upper limit $x_F$.

These two different ways are closely connected: the form parameters are identical $\gamma = \xi$, and as we shall show (see Corollary 1 and Corollary 3 below), if one observes a Poissonian flow of rv characterized with the *GPD*-distribution of exceedance, then the maximum of this Poissonian flow is distributed with the *GEV* distribution. We shall use this duality for the estimation of the tail characteristics of the distribution of earthquake magnitudes for earthquake occurring over arbitrary time interval.

**Definition of the maximum magnitude of an earthquake flow given by a Poisson process :** $M_T$ *is called the T-maximum of a Poissonian flow of observations $X_1, ..., X_v$, generated by the distribution function (DF) $F(x)$, if $M_T = max(X_1, ..., X_v)$ and the following properties are verified: $v$ is a random Poissonian rv with parameter $\lambda T$, which is independent of the $X_j$'s; $\lambda$ is the intensity of the Poissonian flow.*

*The DF of $M_T$ under the condition that one or more observations occur in the interval (0; T) is equal to*

$$\Psi_T(x) = \frac{\exp(-\lambda T[1-F(x)]) - \exp(-\lambda T)}{1 - \exp(-\lambda T)} \cong \exp(-\lambda T[1-F(x)]) \; if \; \lambda T >> 1 \; (Lomnitz \; formula). \quad (7)$$

We state three Corollaries from the *FFT* and *GPBH* theorems.

**Corollary 1.**
*Up to terms of order $\exp(-\lambda T)$, the T-maximum $M_T$ can have a GEV-distribution if and only if the DF $F(x)$ is a GPD-distribution: $F(x) = G_H(x \,|\xi, s) \equiv 1 - [1 + \xi (x-H)/s]^{1/\xi}, \; x \geq H,$ for some $\xi$, $s$ and $H$.*

**Corollary 2.**
*Let $X$ be a rv distributed according to a GPD with DF $G_H(x \,|\xi, s)$ and $K$ is some threshold $K > H$. Then, the conditional distribution of $X$ under the condition $X > K$ is the GPD written as $G_K(x \,|\xi, S)$ with*

$$S = s + \xi(K - H). \quad (8)$$

**Corollary 3.**
*If the rv $X$ has the GPD $G_H(x \,|\xi, s)$ and one observes the maximum $M_T$ (for $\lambda T >> 1$), then $M_T$ has (up to terms of order $\exp(-\lambda T)$) the GEV-distribution $GEV(x \,|\zeta(T), \mu(T), \sigma(T))$ with*

$$\zeta(T) \equiv \xi \; ; \; \sigma(T) = s \cdot (\lambda T)^{\xi} \; ; \; \mu(T) = H - (s/\xi) \cdot [1 - (\lambda T)^{\xi}]. \quad (9)$$

These corollaries are proved in the three Appendix A, B and C.

**Definition of quantiles of the DF of magnitudes**: Once the parameters $\hat{H}$, $\hat{s}$ *and* $\hat{\bar{\xi}}$ have been estimated by one of the methods described below, we can determine any desired statistical characteristic of the tail distribution. One such characteristic, which is both useful and well-behaved, is the quantile $Q_\tau(q)$ at a prescribed confidence level $q$:

$$Q_\tau(q) = \hat{H} - (\hat{s}/\hat{\bar{\xi}}) \cdot [1 - (\frac{\lambda \tau}{\log(1/q)})^{\hat{\bar{\xi}}}]. \quad (10)$$

### 2.3 Two methods for estimating the tail characteristics

In this section, we give the two algorithms based on the two main theorems of EVT that allows us to estimate the distribution of extreme values from a given data set.

#### 2.3.1 The GEV Method

1. Choose an interval of values $(T_L \,; T_H)$ for time interval durations $T$, for which the limit *FFT* and *GPBH* theorems are (approximately) valid and, at the same time, the catalog still contains a sufficient number of $T$-intervals;

2. Choose in this interval $(T_L \,; T_H)$ a finite set of $u$ time-interval durations $T$ $(T_L \leq T_1 < T_2 < ... < T_u \leq T_H)$;

3. Derive the estimates of the *GEV* parameters by the method of moments (found to be more efficient than the ML method for small sample sizes (Pisarenko et al., 2008) for each of the $u$ time interval durations $T$, which yields the following set of parameters:

$\zeta(T_1), \zeta(T_2), ..., \zeta(T_u), \quad \sigma(T_1), \sigma(T_2), ..., \sigma(T_u), \quad \mu(T_1), \mu(T_2), ..., \mu(T_u);$

4.  Using the fact that $\xi = \zeta$ and the relations

$$log(\sigma(T_k)) = log(s) + \xi\, log(\lambda T_k), \qquad k=1,...,u, \qquad (11)$$

$$H = \mu(T_k) + (s/\xi)\cdot[1 - (\lambda T_k)^\xi], \qquad k=1,...,u, \qquad (12)$$

following from equations (9), estimate the average values $\hat{\bar{\xi}}$, $\hat{s}$, $\hat{H}$ of the *GPD* parameters $\xi$, $s$, $H$ by regressing $log(\sigma(T_k))$ and $H$ as a function of $T_k$ for $k=1,..., u$.

5.  Using Lomnitz formula (7), estimate the *DF* of the maximum magnitude $M_\tau$ of a Poissonian flow of main shocks over an arbitrary time interval $\tau$:

$$\Psi_\tau(x) = exp(-\lambda\tau\,[1 + \hat{\bar{\xi}}\,\,(x-\hat{H})/\hat{s}\,]^{-1/\hat{\bar{\xi}}})\,. \qquad (13)$$

### 2.3.2 The GPD Method

1.  Choose an interval $(H_L\,;\,H_H)$ of possible thresholds $H$ for which the limit *FFT* and *GPBH* theorems are (approximately) valid, and at the same time there is a sufficient number of observations over these thresholds;

2.  Choose in this interval $(H_L\,;\,H_H)$ a finite set of $r$ thresholds $H$
    $(H_L \leq H_1 < H_2 < ... < H_r \leq H_H)$;

3.  Derive the estimates of the *GPD* parameters by the *ML* (maximum likelihood) method for each of these $r$ thresholds, which yields
    $\xi(H_1), \xi(H_2),... \xi(H_r);$       $s(H_1), s(H_2),... s(H_r);$            (14)

4.  Using the relations
    $$s(H_k) = s(H_1) + \xi\,(H_k - H_1), \qquad k=2,...,r, \qquad (15)$$
    following from equation (8), estimate the average values $\hat{\bar{\xi}}$, $\hat{s}_1$ of the *GPD* parameters $\xi$, $s(H_1)$ by regressing $s(H_k)$ as a function of $H_k$.

5.  Use Lomnitz formula (7) to get the estimated *DF* of the maximum magnitude $M_\tau$ of a Poissonian flow of main shocks over an arbitrary time interval $\tau$:

$$\Psi_\tau(x) = exp(-\lambda\tau\,[1 + \hat{\bar{\xi}}\,(x - H_1)/\,\hat{s}_1\,]^{-1/\hat{\bar{\xi}}})\,. \qquad (16)$$

### 2.4 GEV versus GPD: pros and cons.

Sections 2.2 and relations (8,9) describe a one-to-one relationship between the two limit distributions for maxima, *GPD* and *GEV*, for sufficiently high threshold $H$ and large time intervals $T$. The two relations (8,9) are very important. They provide the estimates of the *GEV*-parameters for different time window sizes $T$ using regressions based on equations (10,11) and of the GDP parameters for different thresholds $H$ using a regression based on equations (14). The limit distribution of maxima can thus be obtained in two ways: through the *GEV*-approach (when we choose appropriate intervals $T$ and estimate three parameters $\mu, \sigma, \zeta = \xi$) or through the *GPD*-approach (when we choose sufficiently large thresholds $H$ and estimate two parameters $H,\ \xi$),

Which one is better? A priori, each method needs to set one arbitrary value: the time interval duration $T$ in the *GEV* method and the threshold $H$ in the *GPD* method. On the one hand, the *GPD* method requires the determination of only two parameters, compared with the three parameters needed in the *GEV* method. On the other hand, the *GEV* method relies on *T*-maxima (magnitudes in time intervals of duration *T*), which can include magnitudes smaller than the lower threshold *H* used in the *GPD* method. This can reveal more information from the catalog. A contradicting argument is that, if two or more magnitudes larger than *H* are observed in a given *T*-interval, the *GEV* method keeps only the largest one while the *GDP* method will use all of them. It should be remembered that increasing *H* decreases the Poisson intensity $\lambda$ of the flow of events with magnitudes exceeding *H*, thus lowering the product $\lambda T$, which is dangerous for reliable statistical estimations with the *GDP* method. On the other hand, increasing *T* so as to get closer to the asymptotic conditions of application of the FFT theorem decreases the number of *T*-intervals inversely proportionally, leading to poorer statistical estimations with the *GEV* method. Our practical experience with the Harvard global catalog of seismic moments over the period 1977-2006 and with the Fennoscandia region which is presented here shows that the two methods give basically the same efficiency. We thus recommend using both methods simultaneously to check their mutual consistency.

### 2.5 Practical implementation (bootstrap, reshuffling)

In order to implement them, both *GEV* and *GPD* methods can be improved by bootstrap-like procedures that we now describe.

In the *GEV*-approach, this can be done by using the following property of Poissonian flows: if the number of Poissonian events occurring in the interval *(0, T)* is fixed, then these times are distributed as *uniform* iid rv on this interval. As a consequence, starting from the initial catalog, we can reshuffle the occurrence times of all the main shock earthquakes in the catalog $n_s$ times and thus obtain $n_s$ catalog replicas, which should have the same statistical properties. Then, applying step 3 of section 2.3.1 to each of these $n_s$ catalog replicas, we can average over the $n_s$ resulting estimates of the parameters. However, it would be an illusion to believe that making $n_s$ go to infinity would result in making the scatter of the average vanish, as it would happen for *independent* samples by the law of large numbers. Because general bootstrap procedures are performed *using the fixed initial sample*, the scatter does not decrease anymore when $n_s$ becomes of the order of $\cong 50 \div 100$, due to the dependences between the bootstrap replicas. Our numerical experiments on samples with known values of the parameters show that it is sufficient to take $n_s$ $\cong 50 \div 100$, which results in a reduction of the standard deviation of the estimates by *10%* to *30%*. This method was advocated and used in (Pisarenko et al., 2008).

Similarly for the *GPD*-approach, the "classical" bootstrap method can be applied before implementing step 3 (estimation of parameters) of section 2.3.2. The bootstrap proceeds as follows. Starting with a fixed sample of Peaks Over Threshold values, say, magnitudes exceeding *6.4, {M₁,M₂,...,Mᵣ; Mⱼ>=6.4, j=1,...r}*, we pick up randomly *r* times one value from this set, i.e., we allow the choice of the same value several times, corresponding to the so-called "sampling with replacement". Such r successive picks of one value from the set *{M₁,M₂,...,Mᵣ; Mⱼ>=6.4, j=1,...r}* defines one bootstrap sample. By construction, this bootstrap sample has a sample size of *r* values. Repeating this operation $n_b$ times provides $n_b$ bootstrap samples, each of them with a sample size *r* values. For each of the bootstrap sample, we get the statistical estimates of the parameters of the GPD. We then average over these $n_b$ statistical estimates and evaluate the corresponding std as well. This bootstrap procedure provides any desired number $n_b$ of bootstrap samples of observations exceeding a given threshold *H*, and one can average the resulting estimates over the $n_b$ samples. This procedure slightly decreases the scatter of the estimates. But of course, making $n_b$ tend to infinity does not result in decreasing the scatter to zero (as it would be for *independent* samples), since all bootstrap procedures are performed *with the fixed initial sample*. This situation is quite similar to the general bootstrap situation. As in the GEV approach, our numerical experiments on samples with known results show that it is enough to take approximately $n_b \cong 50 \div 100$ to exhaust the gains of this bootstrapping. Further increasing $n_b$ does not lead to any more decrease of the scatter of the estimates.

### 2.6 A remark on the instability of $M_{max}$ versus quantiles $Q_\tau(q)$

In the spirit of (Pisarenko et al., 2008), we would like to stress the instability of the traditional parameter $M_{max}$. It follows from eq.(1) that, for negative form parameters $\xi$ of the *GPD* distribution, the maximum magnitude is given by

$$M_{max} = H - s / \xi .$$

Thus, if $\xi$ is close to zero (say, $\xi \in (- 0.2; - 0.1)$ which corresponds to the range of values found for the global Harvard catalog), then the sample estimates of $M_{max}$ can exhibit large spurious bursts due to random errors occurring in the estimation of the form parameter $\xi$. These spurious outliers are clearly visible in synthetic tests using a perfect sample distributed according to a GPD distribution with a slightly negative form parameter $\xi$. For instance, in a simulation of our estimation procedure on *500* synthetic realizations with true *GPD*-parameters equal to $\xi = -0.1; s = 0.5; H = 4.5; M_{max} = 9.5$, we find that one among the *500* realizations gives the fantastic value of $M_{max} = 17.7!$ On these *500* synthetic realizations, the *16%-50%-84%* quantile estimates for $M_{max} =$ are *(8.25; 9.65; 12.15)*. In contrast, for the quantile $Q_{10}(0.97)$, the *16%-50%-84%* estimates are *(7.03; 7.22; 7.42),* which should be compared with the true quantile is *7.25*.

The stability of the estimation of the quantiles $Q_\tau(q)$ can be understood from the following formula

$$Q_\tau(q) = \mu - (\sigma / \xi) \cdot [1 - (\frac{\lambda \tau}{\log(1/q)})^\xi] ,$$

showing that $Q_\tau(q)$ converges to a ***finite value*** (under the condition that $q < 1$) as $\xi$ goes to zero, namely:

$$Q_\tau(q) \rightarrow \mu + \sigma \cdot \log(\frac{\lambda \tau}{\log(1/q)}), \qquad \xi \rightarrow 0.$$

In contrast, $M_{max} = H - s / \xi$ diverges as $\xi$ goes to zero. This is the source of the instability of the estimation of $M_{max}$ compared with the relative stability of the estimation of $Q_\tau(q)$ for samples characterized by negative $\xi$ values close to zero. This reasoning explains why $Q_\tau(q)$ should be intrinsically more stable than $M_{max}$ for small $\xi$.

Of course, if $q \rightarrow 1$, then $Q_\tau(q) \rightarrow M_{max}$, but this convergence occurs extremely slowly, as the logarithm of the logarithm of the difference $1 - q$. Besides, going from the parameter $M_{max}$ to the $q$-quantile $Q_\tau(q)$ does not give up any useful information, since one can always use $Q_\tau(q)$ for sufficiently large $\tau$ and for $q$ close enough to unity, for which $Q_\tau(q)$ becomes arbitrarily close to $M_{max}$.

### 3. Application of the method to global Harvard catalog 1977-2006.

#### 3.1 Preliminary steps

The analysis of the distributions of magnitudes presented below is performed on catalogs of earthquakes occurring in a given region, time period and magnitude range. For our purposes, we restrict our use of the information in the earthquake catalog to just two values for each earthquake: magnitude $m_i$ and occurrence time $t_i$ . We assume that, for each $i$-th earthquake, its magnitude $m_i$ is drawn from an unknown function $F(m \,|H) = F(m)$ describing the probability density distribution of magnitudes exceeding some lower threshold $H$.

For our analysis, we shall also assume the Poissonian property of the occurrence times $t_i$ of the earthquakes. This assumption is evidently in contradiction with the existence of aftershocks, seismic swarms and other clustered seismic events. We thus restrict our analysis to sub-catalogs obtained in the following way, which constitute better approximations of realizations of Poisson processes. In a first step, we decluster our catalogs using the standard Knopoff-Kagan algorithm [Knopoff et al. 1982]. The algorithm works as follows. We start by identifying the largest event in a catalog, whose time, location and magnitude are denoted as (time $t$, location $(\phi_0, \lambda_0)$ in longitude-latitude, magnitude $m_W$). Then, we exclude all events in the time-space window:

$$(t; \quad t + 10^{-0.31 + 0.46\, m_W});$$
$$R(\phi, \; \lambda; \; \phi_0, \; \lambda_0) \leq \; 10^{-0.85 + 0.46\, m_W} ; \qquad (17)$$

where $R$ is the distance in km between points $(\phi, \lambda)$ and $(\phi_0, \lambda_0)$. The window (17) was taken from (*Knopoff et al.1982*). After the first elimination, we identify the next largest event of the remaining earthquakes (excluding the previous one already accounted for). And we apply the same pruning with the same rule for the space-time window associated with this second largest event among the remaining earthquakes. We iterate until the algorithm stops. This leads to a sequence of events $(m_i, t_i)$, $i = 1,...,r$ , where $t_i$ is close to a Poissonian flow, whose intensity is denoted as $\lambda(H) = \lambda$. *Pisarenko et al. (2008)* have quantified the strong gain in declustering obtained by this method, and have documented quantitatively how the remaining events approximate a Poissonian flow.

Then, we calculate the $T$-maxima of the magnitudes (magnitude maxima in successive time intervals of length $T$). This operation transforms a point process of events occurring at random times into a discrete time random process of $T$-maxima, which is much more convenient for statistical analysis. Besides, this operation improves further the declustering of the catalog, since clusters are usually formed by relatively weak events that are mostly eliminated by keeping only the $T$-maxima in each successive time intervals of length $T$. Our analysis restricts $T$ to take sufficiently large values in order to avoid (with some high probability) the occurrence of empty $T$-intervals, with no value for the maximum magnitude. This restriction is equivalent to the condition $\lambda T >> 1$.

We first study the Harvard catalog, from 01.01.1977 to 16.06.2006, of seismic moments $M$ *(dyne-cm)* transformed into magnitudes $m_W$ by the formula

$$m_W = \frac{2}{3} [log_{10} (M) - 16.1].$$

The next section 4 studies the Fennoscandia catalog covering the time period 01.01.1900 – 31.12.2005.

For the Harvard catalog, we restrict our analysis to earthquake of depth smaller than *70 km* and of magnitudes $m_W$ larger than the lower threshold *5.5*, corresponding to 8102 events. After the application of the Knopoff-Kagan algorithm [Knopoff et al. 1982] described above, we are left with 4193 so-called main shock events suggesting that, according to this declustering method, aftershocks and clustered events constituted about 49.4% of the total set of earthquakes.

The complementary cumulative distribution of the magnitudes of the main shocks is shown in Fig.1. One can observe that the graph is very close to a straight line up to $m_W = 7.7$ (seismic moment $M = 4.5 \cdot 10^{27}$ *ergs*), in agreement with the standard Gutenberg-Richter law, while a significant downward bend occurs beyond that value, which contains the 39 largest events. As discussed previously, e.g. by Pisarenko and Sornette (2003, 2004), this small number of events limits the detailed characterization of the departure from the Gutenberg-Richter law.

### 3.2 Application of the GEV method to the Harvard catalog of main shocks (n = 4193)

We now apply the method described in section 2.3.1 to estimate the parameters of the *GEV*-distribution of the *T*-maxima. We use the method of statistical moments, since this method is slightly more efficient for moderate sample sizes than the maximum likelihood method, as shown previously in *(Pisarenko et al. 2008)*. We calculated these estimates for *T*-intervals in the range (50; 250) days. The numbers $N_T$ of different *T*-intervals in the catalog and the corresponding products $\lambda T$ are the following:

$$
\begin{array}{llllllllll}
T & 50 & 75 & 100 & 125 & 150 & 175 & 200 & 225 & 250 \\
N_T & 214 & 143 & 107 & 85 & 71 & 61 & 53 & 47 & 42 \\
\lambda T & 19.5 & 29.3 & 39.1 & 48.9 & 58.6 & 68.4 & 78.2 & 87.9 & 97.7
\end{array}
\tag{18}
$$

The Poisson intensity $\lambda$ is determined as *n/number of days = 4193/10728 = 0.3908*. All *T*-intervals are non-empty. For *T*-values exceeding 150 days, the number $N_T$ of different *T*-intervals in the catalog is too small for a reliable estimation of the three *GEV*-parameters. Figs.2-4 show the moment-estimates of the *GEV*-parameters for the declustered Harvard catalog. Fig.2 shows an approximate "stabilization" of the ξ-estimates in the range $75 \le T \le 175$. The scale and location estimates of $\sigma(T)$ and $\mu(T)$ do not provide any useful restriction on the usable values of *T*, as they behave as prescribed by equation (9).

In order to choose an appropriate interval for the *T*-values, we consider the Kolmogorov distances between the sample distribution of the *T*-maxima and the fitted *GEV*-distribution function. The Kolmogorov distance *KD* is defined as follows:

$$
KD = n_T^{1/2} max \mid \hat{F}_T(x) - \Phi_T(x \mid \bar{\mu}, \bar{\sigma}, \hat{\bar{\xi}}\,) \mid,
\tag{19}
$$

where $(\bar{\mu}, \bar{\sigma}, \hat{\bar{\xi}})$ are the *ML*-estimates of the *GEV*-parameters, $\hat{F}_T(x)$ is the sample cumulative distribution of the *T*-maxima. Since we use a theoretical function $\Phi_T$ with parameters fitted on the data, we cannot use the standard Kolmogorov distribution to check the statistical significance of the sample value of *KD*. Instead, in order to determine the confidence level of a given *KD*-distance obtained for a given sample, one has to use a numerically simulated distribution of *KD*-distances in the simulation procedure using random *GEV* samples with the fitted parameters $(\bar{\mu}, \bar{\sigma}, \hat{\bar{\xi}})$. The artificial *GEV* samples simulating our real sample were taken with the following parameters

$T = 80$ days;
$\sigma(T) = s \cdot (\lambda T)^{\bar{\xi}} = 0.84 \ (31.3)^{-0.185} = 0.45;$
$\mu(T) = H - (s/\bar{\xi}) \cdot [1 - (\lambda T)^{\bar{\xi}}] = 4.98 + (0.84/0.185) \cdot [1 - (31.3)^{-0.185}\ ] = 7.12;$
$\zeta = \bar{\xi} = -0.185; \quad H = 4.98; \quad s = 0.85$ .
$$\tag{20}$$

These values are close to our estimates for the declustered Harvard catalog. We simulated our estimation procedure *1000* times and obtained numerically the probability

$$
p(z) = P\{\ KD \ge z \mid \bar{\xi} = -0.185;\ s = 0.85;\ h = 4.98\}.
$$

The obtained probabilities for different values of the argument *z* are
$$
p(0.55) = 0.53;\ p(0.6) = 0.30;\ p(0.65) = 0.17;\ p(0.7) = 0.073;\ p(0.75) = 0.027;
\tag{21}
$$

We can thus consider a value $KD \leq 0.65$ as acceptable, while $KD \geq 0.7$ is taken as contradicting the GEV-distribution. Fig.5 shows that we should therefore exclude from our estimation procedure $T$-intervals with $T < 50$ days.

Combining the evidence for an approximate "stabilization" of the $\xi$-estimates in the range $75 \leq T \leq 175$ and the criterion provided by the $KD$ statistics, we decide to choose the range $75 \leq T \leq 150$ for the $T$-intervals, in discrete values $T = 75; 100; 125; 150$, to apply equations (8,9) for the estimates of $\xi, \sigma, \mu$. We used $100$ shuffled samples to decrease the scatter of these regression estimates. As final estimates, we suggest to take the medians of the corresponding $100$ estimates. We show below these final estimates together with the 16%- and 84%-quantiles, characterizing the bootstrap scatter:

| quantile level | *16%* | median *(50%)* | *84%* | |
|---|---|---|---|---|
| $\xi$ | *-0.221* | ***-0.185*** | *-0.152* | (22) |
| $s$ | *0.714* | ***0.847*** | *1.025* | |
| $H$ | *4.669* | ***4.982*** | *5.236* | |
| $M_{max}$ | *9.32* | ***9.58*** | *9.96* | |
| $Q_{10}(0.97)$ | *8.96* | ***9.06*** | *9.16* | |

As we shall see below, the estimates of the form parameter $\xi$ and of the tail variables $M_{max}$ and $Q_{10}(0.97)$ are compatible with those obtained below in section 3.3 using the GPD method. However, we will see that the estimates of the other two parameters $\sigma, \mu$ differ rather significantly between the two methods, and we provide the explanation of this discrepancy.

In order to estimate the Mean Square Error (*MSE*) of these estimates, we implemented our whole estimation procedure on $N=500$ artificially generated *GEV* samples with the following parameters: $\xi = -0.185; \sigma = 0.45; \mu = 7.12; n = 4193; \lambda = 0.3908; T = 80$ *days; $N_s$=100.* We obtained the following results:

$$MSE(\xi) = 0.047; \quad MSE(\sigma) = 0.145; \quad MSE(M_{max}) = 0.68; \quad MSE(Q_{10}(0.97)) = 0.23 \qquad (23)$$

It should be remarked that in all cases the bias was much smaller than the Std for all the parameters. We stress once more than the scatter of $M_{max}$ is much larger than that of $Q_{10}(0.97)$. The values reported in (23) can be taken as estimates of the *real scatter*. This "real" scatter is different from and larger than that obtained with the bootstrap procedure. This is not surprising since the latter gives only the scatter conditional to the same unique data sample. Thus, our final results for the estimation of the GEV parameters by the GEV method can be summarized as:

$$\hat{\xi} = -0.185 \pm 0.047; \qquad (24)$$

$$\hat{s} = 0.847 \pm 0.145;$$

$$\hat{H} = 4.982 \pm 0.198;$$

$$\hat{M}_{max} = 9.58 \pm 0.68;$$

$$\hat{Q}_{10}(0.97) = 9.06 \pm 0.23.$$

### 3.3 Application of the GPD method to the Harvard catalog of main shocks *(n = 4193)*

We now apply the method described in section 2.3.2 to estimate the parameters of the *GPD*-distribution of the earthquake magnitudes in the declustered Harvard catalog above thresholds $H$ chosen in the interval *(6.0; 7.6)*. The set of thresholds and corresponding numbers $n_H$ of exceedances were the following:

| $H$ | 6.0 | 6.2 | 6.4 | 6.6 | 6.8 | 7.0 | 7.2 | 7.4 | 7.6 | |
|---|---|---|---|---|---|---|---|---|---|---|
| $n_H$ | 1454 | 941 | 625 | 392 | 264 | 174 | 114 | 80 | 50 | (25) |

Thresholds larger than *7.4* are hardly admissible, since the number of exceedances becomes too small.

Figs. 6 and 7 plot the estimates of the parameters $\xi$, and $s$ obtained by the maximum likelihood method with $N_b = 100$ bootstrap samples. Equation (15) predicts that, if the form parameter $\xi$ is negative, the $s$-estimates should decrease with the threshold $H$. However, one can observe in Fig. 7 that the negative $\xi$-estimates shown in Fig.6 are associated with decreasing $s$-estimates only for $m_W \geq 6.6$. One can thus draw the conclusion that threshold magnitudes smaller than *6.6* are too small for a proper application of the limit *GBPH* theorem. For our following analysis, we thus impose the restriction that all thresholds will be taken larger than *6.6*.

Fig.6 exhibits a *stabilization* (approximate constant value) of the estimates of the form parameter $\xi$ (although the scatter for $h > 7.2$ become very large), in the interval $6.6 \leq H \leq 7.4$. This stabilization suggests that the interval $6.6 \leq H \leq 7.4$ for the estimation of the *GPD* parameters is appropriate.

The Kolmogorov distances between the sample excess functions and the fitted *GPD*-distribution function provides a systematic approach to determine the appropriate interval for the thresholds. In the present application, the Kolmogorov distance $KD$ is defined as follows:

$$KD = n_H^{1/2} max \mid \hat{F}_H(x) - G_H(x \mid \hat{\xi}, \hat{s})\mid, \qquad (26)$$

where $(\hat{\xi}, \hat{s})$ are the *ML*-estimates of the *GPD*-parameters, and $\hat{F}_H(x)$ is the sample stepwise excess function. Since we use a theoretical function $G_H$ with parameters fitted on the data, we cannot use the standard Kolmogorov distribution to check the sample value of the *KD*. In order to determine the confidence level of a given *KD*-distance obtained for a given sample, one has to use a numerically simulated distribution of *KD*-distances in the simulation procedure using random *GPD* samples with the same fitted parameters $(\hat{\xi}, \hat{s})$. Fig.8 shows the *KD*-distances for thresholds $H$ in the interval $6.4 \leq H \leq 7.6$. Artificial *GPD* samples simulating our real sample were taken with parameters $\xi = -0.2$; $s = 0.53$ close to the values obtained from the empirical data. We simulated our estimation procedure *1000* times and estimated numerically the probability $p(z) = P\{ KD \geq z \mid \xi = -0.2; s = 0.53\}$. We obtained the following values:

$$p(0.9) = 0.52; \ p(1.0) = 0.27; \ p(1.1) = 0.08; \ p(1.2) = 0.04. \qquad (27)$$

This leads us to consider $KD \leq 1.0$ as acceptable, while $KD \geq 1.1$ as contradicting the hypothesis that the data is generated by the *GPD*-distribution. We thus exclude values of thresholds $H$ smaller than *6.5*.

Combining these different conditions into the regressions used to derive the estimates of $\xi$, and $s_1 = s(h_1)$ (with equation (15)), we decided to restrict the range of adequate thresholds to the interval $6.6 \leq H \leq 7.2$, sampled with the following discrete values: $h = 6.6; 6.8; 7.0; 7.2$. We used 100 bootstrap samples so as to decrease the scatter of these regression estimates. As final estimates, we suggest to consider the medians of these 100 estimates. The later are shown below together with the 16%- and 84%-quantiles used to characterize the bootstrap scatter:

| quantile level | *16%* | median *(50%)* | *84%* | |
|---|---|---|---|---|
| $\xi$ | *-0.350* | ***-0.204*** | *-0.148* | (28) |
| $s$ | *0.481* | ***0.529*** | *0.584* | |

| | | | |
|---|---|---|---|
| $M_{max}$ | 8.63 | **9.53** | 10.31 |
| $Q_{10}(0.97)$ | 9.196 | **9.212** | 9.226 |

Note that the distribution of ξ-estimates is rather asymmetrical, and that the scatter of $M_{max}$ is significantly larger than that of $Q_{10}(0.97)$. In order to estimate the Mean Square Error (*MSE*) of these estimates given by $MSE^2 = Bias^2 + Std^2$, we simulated *N=500* times our full estimation procedure for artificially generated *GPD* samples with parameters: $ξ = -0.2$; $s = 0.53$; $H = 6.6$; $n_H = 293$; $N_b = 100$ and obtained the following results:

$$MSE(ξ) = 0.049; \quad MSE(s) = 0.039; \quad MSE(M_{max}) = 0.50; \quad MSE(Q_{10}(0.97)) = 0.20 \qquad (29)$$

In all cases, the bias was found much smaller than the Std for all parameter values. The values reported in (29) can be taken as reasonable estimates of the *real scatter* (in contrast to the conditional scatter obtained from the bootstrap procedure). Thus, our final results on the estimation of the *GPD* parameters by the *GPD* method are

$$\hat{ξ} = -0.204 \pm 0.049;$$
$$\hat{s} = 0.529 \pm 0.039;$$
$$\hat{M}_{max} = 9.53 \pm 0.52; \qquad (30)$$
$$\hat{Q}_{10}(0.97) = 9.21 \pm 0.20.$$

### 3.4 Comparison between the results of the GEV and GPD methods

We now discuss the consistency of the estimations of the form parameter ξ, of the scale parameter *s* and of the tail parameters $M_{max}$ and $Q_{10}(0.97)$, obtained respectively by the *GEV* and *GPD* methods.

Let us first address the large apparent discrepancy between the scale parameter estimate *0.847* in eq.(22) obtained by the *GEV* and the corresponding estimate *0.529* in eq.(28) obtained by the *GPD* method. It turns out that there is a simple explanation for this discrepancy, which has to do with (i) the fact that the parameters obtained with the *GEV* method imply a lower threshold which is different from the threshold used in the GPD method and (ii) the dependence of the scale factor on the value of the used threshold. To show this, let us recall that the correspondence stated in the Corollaries 1 and 3 of section 2.2 between the *GEV* and *GPD* distributions asserts that

$$exp\{ -[1 + ξ(x-μ)/σ]^{-1/ξ} \} = exp(-λT[1-F(x)]), \quad x \le μ - σ/ξ; \quad (λT >> 1), \qquad (31)$$

where *F(x)* is the *DF* of the observed magnitudes. It follows from eq.(31) that, up to terms of order *exp(-λT)*, we have

$$F(x) = 1 - \frac{1}{λT}[1 + \frac{ξ}{σ}(x - μ)]^{-1/ξ}, \qquad (32)$$

which shows that the *DF F(x)* has the form of a *GPD* distribution. Since *F(x)* is non-negative, we have to restrict the domain of definition of *F(x)* from below, that is, *F(x)* approximated by (32) is found non-negative for $μ + \frac{σ}{ξ}[(λT)^{-ξ} - 1] \le x$. Since the upper bound $x \le μ - σ/ξ$ is also imposed from (31), this implies that, up to terms of order *exp(-λT)*, the DF *F(x)* varies from zero to approximately unity over the interval

$$\mu + \frac{\sigma}{\xi} [ \ (\lambda T)^{-\xi} - 1] \leq x \leq \ \mu - \sigma/\xi \qquad (33)$$

However, there is no reason for the left boundary in (33) to be equal to the threshold $H$ used in the *GPD* method. Indeed, if we accept for $\xi$, $\sigma$, $\mu$ the values given by equation (20), we get the lower boundary of the interval defined in (33) equal to $\mu + \frac{\sigma}{\xi} [ \ (\lambda T)^{-\xi} - 1] \cong 5.0$. This is very different from the lower threshold $H = 6.6$ used in the *GPD* method. But Corollary 2 shows that the *GPD*-distribution has its scale parameter changed under a shifting threshold according to the formula $S = s + \xi \ (H - K)$. If we take $H - K = 6.6 - 5.0 = 1.6$; $\xi = -0.185$; $s = 0.847$ (see eq.(24)), then we get $S = 0.847 - 0.185 \cdot 1.6 = 0.551$ which is close to the $s$-estimate in eq.(28) obtained by the *GPD* method. We can thus conclude that the values *0.847* (eq.(22) obtained by the *GEV*) and *0.529* (eq.(28) obtained by the *GPD* method) are actually compatible, when adjusted to the same effective threshold $H$.

Concerning the form parameter $\xi$, both methods give very similar results $\hat{\xi} = \mathbf{-0.185 \pm 0.047}$ (eq.(24) for the *GEV* method) and $\hat{\xi} = \mathbf{-0.204 \pm 0.049}$ (eq.(30) for the *GPD* method), with basically the same scatter.

The estimates of $\hat{M}_{max} = \mathbf{9.58 \pm 0.68}$ for the *GEV* method and $\hat{M}_{max} = \mathbf{9.53 \pm 0.52}$ for the *GPD* method are highly consistent; similarly for the quantile $\hat{Q}_{10}(0.97) = \mathbf{9.06 \pm 0.23}$ for the *GEV* method and $\hat{Q}_{10}(0.97) = \mathbf{9.21 \pm 0.20}$ for the *GPD* method. As the two methods exploit the data in complementary ways and appear to be similarly efficient, we recommend using both of them simultaneously to check their consistency on unknown data sets. These results as well as additional synthetic tests confirm that our methods provide efficient ways to recover the correct parameters. However, while there is no significant bias, the uncertainty on $\hat{M}_{max}$ is about two to three times larger than that on $\hat{Q}_{10}(0.97)$. In addition, the distribution of $\hat{M}_{max}$ exhibits a fat tail towards large values, making its determination uncertain.

### 4. Estimation of GEV and GPD parameters for discrete magnitudes. Application to Fennoscandia (01.01.1900 – 31.12.2005)

#### 4.1 Preliminary considerations

In many regions, magnitude catalogs report magnitudes with a discrete scale. This is mainly due to human truncation biases associated with the use of Intensity scales converted to magnitude scales. We thus investigate how to adapt our methods to this situation.

As a representative example, we consider the catalog of the Fennoscandia [Ahjes and Uski, 1992; Uski and Pelkonen, 2006] covering the time period 01.01.1900 – 31.12.2005 and the space domain restricted by polygons with the coordinates:

| Latitude | Longitude |
|----------|-----------|
| *72.0* | *20.0* |
| *72.0* | *40.0* |
| *70.0* | *40.0* |
| *50.0* | *10.0* |
| 55.0 | *0* |
| *60.0* | *-5.0* |
| 65.0 | *0* |

A total number of *8381* events with magnitudes *-0.7 ≤ m ≤ 5.8;* depth *0 ≤ H ≤ 97 km* occurred in this space-time interval. The event with maximum magnitude occurred on *01.09.1819* with coordinates *λ = 66.4; φ = 14.4;* depth unknown (Norway).

The seismic flow exhibits some variability. Fig.9 shows the yearly average number of shocks in moving windows of *50* years duration for four lower thresholds: *m = -0.7 (n = 7298 events); m = 3.0 (n = 964 events); m = 3.5 (n = 450 events); m = 4.0 (n = 197 events).* For the lowest threshold *m = -0.7,* one can observe a positive trend taking off around 1940-2005, contrasted by a negative trend before 1940. For higher thresholds *(m = 3.0; 3.5; 4.0),* the positive and negative oscillations are comparable, and no clear tendency prevails on the whole time interval 1900-2005. Thus, we can conclude that

1. On the time interval 1900-2005, the seismic flow *m ≥ 3.0 (n=964)* can be considered as approximately stationary;
2. A more cautionary approach to ensure a better stationarity of the seismic flow would consists in taking the lower threshold *m = 3.5,* although the corresponding sample size becomes rather low *(n=450).*

Since we are interested in the analysis of main shocks, we have applied the algorithm developed by Knopoff-Kagan to remove a significant part of the aftershocks from the catalog. We obtained the following numbers of main shocks:

*m = -0.7      6868* main shocks;    intensity *λ =0.177* events/day = *64.7* events/year
*m = 3.0      928* main shocks;    intensity *λ =0.0240* events/day = *8.75* events/year
*m = 3.5      433* main shocks;    intensity *λ =0.0112* events/day = *4.08* events/year
*m = 4.0      190* main shocks;    intensity *λ =0.0049* events/day = *1.79* events/year

The intensity of the seismic flow restricted to the set of main shocks for these 4 thresholds is paralleling very closely that shown in fig. 9 for all earthquakes. It should be noted that the percentage of aftershocks is very low for Fennoscandia, about 6%, compared to other seismic zones, where it reaches 50% and more.

Fig.10 shows the histogram of the main shock magnitudes. The discreteness of the reported magnitudes is clearly visible. The histogram becomes smoother and regularly decreasing for *m ≥ 2.2.* There is some slight evidence of a preeminence of half-integer magnitude values, but the effect is not large. This leads to conclude that we can apply our methods to the main shocks in the magnitude range *m ≥ 3.0,* which ensures an approximate stationary seismic flow and a reasonably smooth distribution. Fig. 11 shows the sample tail *1 – F(x)* of the main shock magnitudes (*1900-2005*). For *m ≥ 2.2,* an approximate linear dependence can be observed, followed beyond *m = 4.0* by a somewhat steeper slope, which is indicative of a faster decay of the magnitude *PDF* in this range.

We now apply the *GEV* and *GPD* methods to this sample of main shocks with magnitudes *m ≥ 3.0,* *n = 928.* The magnitudes in the catalog are quantized with 0.1 units. First of all, let us show that this quantization does not affect significantly the resulting estimates. We take just one illustrative example, namely synthetic samples generated with a *GPD*-distribution with parameters close to the estimates determined below for the real catalog: $\xi = -0.275;$ s = 0.776. For a lower threshold *H = 2.5,* we took a sample size *n = 2200* (which is close to the value for the Fennoscandia catalog). We use the *GPD* method of estimation with $N_b$ bootstraps. We quantized our sample of magnitudes in step *dm = 0.1; 0.15; 0.2; 0.25* and then compared the estimates obtained with the continuous and discrete samples. We found that, for *dm = 0.1* and even perhaps *dm = 0.15,* the mean square difference of the estimates for discretely and continuously sampled magnitudes is negligible compared with the std of the estimates. For *dm = 0.2* and larger, the effect of quantization should be taken into account. These tests are summarized in the following table. Denoting the Mean Square Difference of the estimates obtained with the continuous and discrete magnitude sampling by *MSD(ξ)* and *MSD(s)*, we have:

| *dm* | *0.10* | *0.15* | *0.20* | *0.25* |
|---|---|---|---|---|
| *MSD(ξ)/std(ξ)* | *0.14* | *0.26* | *0.29* | *0.50* |
| *MSD(s)/std(s)* | *0.11* | *0.23* | *0.28* | *0.56* |

Our conclusion about neglecting the quantization effect for *dm = 0.1*, which corresponds to the case of the Fennoscandia catalog, can be considered as justified.

For the sake of completeness, let us briefly point out the modifications that would be needed for data with larger magnitude steps. One can indeed account exactly for discrete magnitudes by using the discrete analog of the Likelihood $L^d$ (for the *GPD* approach):

$$L^d = (n_1, ..., n_r \mid H, s, s\xi) = \prod_{k=1}^{r} P_k(n_k \mid H, s, \xi).$$

Here $n_k$ is the number of occurrences of a given discrete magnitude $m_k$, and $P_k$ are the probabilities calculated from the *GPD*-distribution:

$$P_k(n_k \mid H, s, \xi) = G_H(m+k\Delta m \mid s,\xi) - G_H(m+(k-1)\Delta m \mid s,\xi), \ k=1,2,...,r.$$

The discrete likelihood $L^d$ must be maximized over the parameters *(s,ξ)* by numerical methods. For the *GEV* approach, we would recommend to use the so-called Sheppard's corrections (see *(Cramer 1940)*) to statistical moments calculated from discrete data.

### 4.2 Application of the GEV method to the Fennoscandia catalog of main shocks (m > 3.0, n = 928)

We calculated the estimates of the *GEV* distribution for *T*-intervals in the range *(100; 1500)* days. The numbers $N_T$ of different *T*-intervals in the catalog and the corresponding products $\lambda T$ are the following:

| *T* | *100* | *300* | *500* | *700* | *900* | *1100* | *1300* | *1500* | |
|---|---|---|---|---|---|---|---|---|---|
| $N_T$ | *387* | *129* | *77* | *55* | *43* | *35* | *29* | *25* | (34) |
| $\lambda T$ | *2.40* | *7.19* | *12.0* | *16.8* | *21.6* | *26.4* | *31.2* | *36.0* | |

The Poisson intensity $\lambda$ was determined as *n/number of days = 928/38716 = 0.0240*. All *T*-intervals larger than *T = 300* are found non-empty. *T*-values exceeding *900* days should not be used in the estimation procedure, as the number of different *T*-intervals in the catalog becomes too small for a reliable estimation of the three *GEV*-parameters. Figs. 12-14 show the moment-estimates of the three *GEV*-parameters. Fig.12 exhibits an approximate plateau for the dependence of the ξ-estimates as a function of *T*, in the range $300 \leq T \leq 900$. The estimates of $\sigma(T)$ and $\mu(T)$ shown respectively in Figs. 13 and 14 do not impose any restriction on the values of *T*, since they behave as prescribed by equations (11,12).

The calculations of the *KD*-distances defined by expression (19) following the method of section 3.2 show that we should exclude windows with $T \leq 400$ days. For the calculations of the *KD*-distances, we used artificial *GEV* samples simulating our real data with parameters *T = 400* days; $\sigma(T) = s \cdot (\lambda T)^\xi = 0.36$; $\mu(T) = H - (s/\xi) \cdot [1 - (\lambda T)^\xi] = 4.05$; $\zeta = \xi = -0.275$; *H=2.93*; *s=0.67*. We simulated our estimation procedure *1000* times and estimated numerically the probability $p(z) = P\{ KD \geq z \mid \xi = -0.275; s = 0.67; H = 2.93\}$. We obtained *p(0.55)= 0.52; p(0.6) = 0.27; p(0.65) = 0.15; p(0.7) = 0.05; p(0.75) = 0.02*, showing that values $KD \leq 0.65$ are acceptable, while values of $KD \geq 0.7$ reject at the *95%* confidence level the hypothesis that the data could be generated with the *GEV*-distribution.

Combining these different constraints, we decided to keep for the analysis and for the regression estimates of $\xi, s, H$ the *T*-intervals with *T* in the range $400 \leq T \leq 800$, sampled with the following discrete values: $T = 400; 500; 600; 700; 800$. We used *100* shuffled samples in order to decrease the scatter of these regression estimates. As final estimates, we suggest to take the medians of the corresponding *100* estimates. We show below these final estimates together with the *16%-* and *84%*-quantiles, which are characterizing the scatter of the bootstrap procedure:

| quantile level | *16%* | median *(50%)* | *84%* | |
|---|---|---|---|---|
| $\zeta$ | *-0.315* | ***-0.262*** | *-0.213* | (35) |
| $s$ | *0.631* | ***0.763*** | *0.937* | |
| $H$ | *2.70* | ***2.94*** | *3.13* | |
| $M_{max}$ | *5.68* | ***5.86*** | *6.08* | |
| $Q_{10}(0.97)$ | *5.36* | ***5.41*** | *5.51* | |

In order to evaluate the Mean Square Error (*MSE*) of these estimates, we simulated *N=500* times our whole estimation procedure on artificially generated *GEV* samples with parameters $\zeta = -0.275;\ \sigma = 0.36;\ \mu = 4.05;\ n = 928;\ \lambda = 0.024;\ T = 400\ days;\ N_s = 100$ and obtained the following results:

$$MSE(\zeta) = 0.0434;\ MSE(s) = 0.0769;\ MSE(M_{max}) = 0.211;\ MSE(\ Q_{10}(0.97)\ ) = 0.091 \quad (36)$$

The biases are much smaller than the Std for all parameters. We stress once more than the scatter of $M_{max}$ is much larger than that of $Q_{10}(0.97)$. The *MSE* (36) can be taken as estimates of the *real scatter* (in contrast to the conditional scatter of the bootstrap). Thus, our final results on the estimation of the parameters by the *GEV* method can be summarized as:

$$\hat{\xi} = \textbf{\textit{-0.262} } \pm \textbf{\textit{0.043;}} \tag{37}$$

$$\hat{s} = \textbf{\textit{0.763} } \pm \textbf{\textit{0.077;}}$$

$$\hat{H} = \textbf{\textit{2.94} } \pm \textbf{\textit{0.071;}}$$

$$\hat{M}_{max} = \textbf{\textit{5.86} } \pm \textbf{\textit{0.21;}}$$

$$\textbf{\textit{Q}}_{10}\textbf{\textit{(0.97)}} = \textbf{\textit{5.41} } \pm \textbf{\textit{0.09.}}$$

### 4.3 Application of the GPD method to the Fennoscandia catalog of main shocks (m > 3.0, n = 928)

The *ML*-estimates of the parameters $\xi, s, M_{max}$ are shown in Fig.15-17 as functions of the lower magnitude threshold *H*. We see in Fig.16 that the *s*-estimates decrease as it is requested from eq.(9) only for $H > 3.0$. Thus, thresholds $H \leq 3.0$ are excluded from our analysis. An approximate plateau for the $\xi$-estimates is observed in the range $3.1 \leq H \leq 3.6$. The numbers of exceedences $n_h$ decreases from *928* for *H=3.0* to *190* for *H=4.0*, which is sufficient for our analysis over the whole range $3.0 \leq H \leq 4.0$. We also calculated the *KD*-distances to help constrain further the range of admissible thresholds *H*. Artificial *GPD* samples simulating our real sample were taken with parameters $\xi = -0.275;\ s = 0.67$, close to the estimates using the real data. We simulated our whole estimation procedure *1000* times and evaluated numerically the probability $p(z) = P\{\ KD \geq z\ |\xi = -0.275;\ s = 0.67\}$. We obtained $p(0.9) = 0.58;\ p(1.0) = 0.31;\ p(1.1) = 0.15;\ p(1.2) = 0.07$, showing that $KD \leq 1.0$ can be considered acceptable, while $KD \geq 1.2$ is rejecting the hypothesis that the data is generated with the *GPD*-distribution. This leads to exclude threshold values less than *3.0*. Combining all above mentioned restrictions, we decided to choose the thresholds in the interval $3.05 \leq H \leq 3.75$, sampled with the following discrete values $h = 3.05$ ; $3.15$; $3.25$; $3.35$; $3.45$; $3.55$; $3.65$; $3.75$. We used *100* bootstrap samples to decrease the scatter of the regression estimates $\xi$, $s_1 = s(h_1)$ via equation (15). Following our above analyses, we use the median of the 100 estimates as the final estimates of the parameters and represent their bootstrap scatter with the 16%- and 84%-quantiles:

| quantile level | *16%* | median *(50%)* | *84%* | |
|---|---|---|---|---|
| $\xi$ | *-0.312* | ***-0.278*** | *-0.245* | (38) |
| $s$ | *0.632* | ***0.669*** | *0.706* | |
| $M_{max}$ | *5.60* | ***5.76*** | *5.96* | |
| $Q_{10}(0.97)$ | *5.35* | ***5.44*** | *5.54* | |

The distribution of the $\xi$-estimates is asymmetrical, and the scatter of $M_{max}$ is again larger than that of $Q_{10}(0.97)$. As in the previous calculations, in order to evaluate the Mean Square Error (*MSE*) of these estimates, we simulated *N=500* times our whole estimation procedure on artificially generated *GPD* samples with parameters $\xi = -0.275$; $s = 0.67$; $H = 3.05$; $n_H = 928$; $N_b = 100$. We obtained the following results:

$$MSE(\xi) = 0.0254; \quad MSE(s) = 0.0294; \quad MSE(M_{max}) = 0.165; \quad MSE(Q_{10}(0.97)) = 0.073 \quad (39)$$

As in previous examples, the biases are much smaller than the Std for all the parameters. The *MSE* given by (39) provide reasonable estimates of the *real scatter*. Thus, our final results for the estimation of the parameters by the *GPD* method can be summarized by the following numbers:

$$\hat{\xi} = \textbf{\textit{-0.275} ± \textit{0.025}};$$
$$\hat{s} = \textbf{\textit{0.669} ± \textit{0.029}};$$
$$\hat{M}_{max} = \textbf{\textit{5.76} ± \textit{0.165}};$$
$$\hat{Q}_{10}(0.97) = \textbf{\textit{5.44} ± \textit{0.073}}.$$

(40)

These estimates are completely consistent with those obtained with the *GEV* method. This can be checked directly for the form parameter $\xi$ and for the tail parameters $M_{max}$, $Q_{10}(0.97)$. The estimates of the parameters *s,H* differ, but this discrepancy results only from using different thresholds, as explained for the Harvard catalog in section 3.4. For the Fennoscandia catalog, the reasoning is the same and the results are parallel to those obtained for the Harvard catalog. We do not repeat the detailed calculation.

## 5. Conclusion

We have applied a new approach combining the two main theorems of EVT to estimate the parameters quantifying the tails of the distribution of large earthquakes in the global Harvard catalog and in the local Fennoscandia catalog. We have developed several procedures and check points to decrease the scatter of the estimates and to verify their consistency. The results are satisfactory and can be used as reasonable estimates both for scientific applications and for risk assessment. We expect that this methodology can be fruitfully applied to many other catalogs, by providing both checks of the quality and reliability of the catalogs and useful estimates of the large seismic risks in terms of both maximum possible magnitudes and quantiles.

The estimation of characteristics of the tail of the earthquake magnitude distribution beyond the range of magnitudes available in the historical record, i.e. for a probability level *q > 1- 1/n* where *n* is sample size, is only possible if some additional assumptions about the distribution function are imposed. Sometimes, such assumptions can be made on physical grounds. In the present paper, such additional assumptions have been formulated on the basis of general limit theorems for maximum values of iid sequence of random values. Of course, there is no a-priori guarantee that such assumptions will be fulfilled in a real situation. In our case, these assumptions boil down to assuming a regular behavior of the tail $1 - F(x)$ in the vicinity of the right limit point of the distribution. The fact that there is no limit theorem without such regular behavior serves as a certain justification of such an assumption. But strictly speaking, it is not verifiable in practice. Extreme Value Theory offers a methodology to extrapolate outside the range of the available data. The question of whether the conditions of $EVT$ are satisfied in a given real problem should be solved in each specific case. As Richard Smith [Smith 1990] said: "But what $EVT$ is doing is making the best use of whatever data you have about extreme phenomena".

*Appendix. Proofs of the three Corollaries.*

*Corollary 1.*

Let $F_H(x)$ be the *GPD*-distribution

$$F_H(x) = G_H(x \,|\, \xi, s) \;=\; 1 - (1 + \xi\,(x-H)/s)^{-1/\xi} \quad , \; x \geq H \qquad (A1)$$

In accordance with Lomnitz formula (7), up to terms of order $exp(-\lambda T)$, the distribution function of the *T*-maxima $M_T$ is given by

$$\Psi_T(x) = exp(-\lambda T\cdot(1 + \xi(x-H)/s)^{-1/\xi}) \; \; if \; \lambda T >> 1. \qquad (A2)$$

If we set

$$\sigma = \sigma(T) = s\cdot(\lambda T); \qquad \mu = \mu(T,H) = H - (s/\xi)[1 - (\lambda T)^{\xi}] \qquad (A3)$$

then we can rewrite (A2) in the form of a *GEV*-distribution

$$\Psi_T(x) = exp\{- [1 + \xi(x-\mu)/\sigma]^{-1/\xi} \}. \qquad (A4)$$

It should be noted that, in equation (A1), $F_H(x)$ is defined only for $x \geq H$, while $\Psi_T(x)$ does not vanish at $x = H$, since:

$$\Psi_T(H) = exp(-\lambda T).$$

But according to the condition of Corollary 1, we can neglect terms of order $exp(-\lambda T)$. Therefore, one can complement the domain of definition of $\Psi_T(H)$ for $x < H$ (as it is required for the *GEV*-distribution), since $\Psi_T(x)$ remains smaller than $exp(-\lambda T)$ for $x < H$.

Inversely, if we assume that $M_T$ have a *GEV*-distribution, then using the transformation law (A3) for the parameters, we have

$$s = \sigma /\cdot(\lambda T); \qquad H = \mu + (s/\xi)[1 - (\lambda T)^{\xi}] \;, \qquad (A5)$$

and we get the distribution function of $M_T$ in the form (A1), from which it follows that $F_H(x)$ is a *GPD*-distribution.

*Corollary 2.*

Let *X* be distributed according to the *GPD*-distribution:

$$F_H(x) = \; 1 - (1 + \xi\,(x-H)/s)^{-1/\xi} \quad , \; x \geq H \,. \qquad (A6)$$

Then, for any $K > H$ the conditional distribution of *X* under the condition $X > K$ is

$$F_K(x) = \{ F_H(x) - F_H(K)\} /\{ 1 - F_h(K)\}, \qquad x \geq K. \qquad (A7)$$

Putting (A6) into (A7), we get

$$F_K(x) = 1 - (1 + \xi\,(x-K)/S)^{-1/\xi} \quad , \; x \geq K \,, \qquad (A8)$$

where

$$S = s + \; \xi\,(K-H). \qquad (A9)$$

*Corollary 3.*

This corollary follows from the proof of Corollary 1 above.

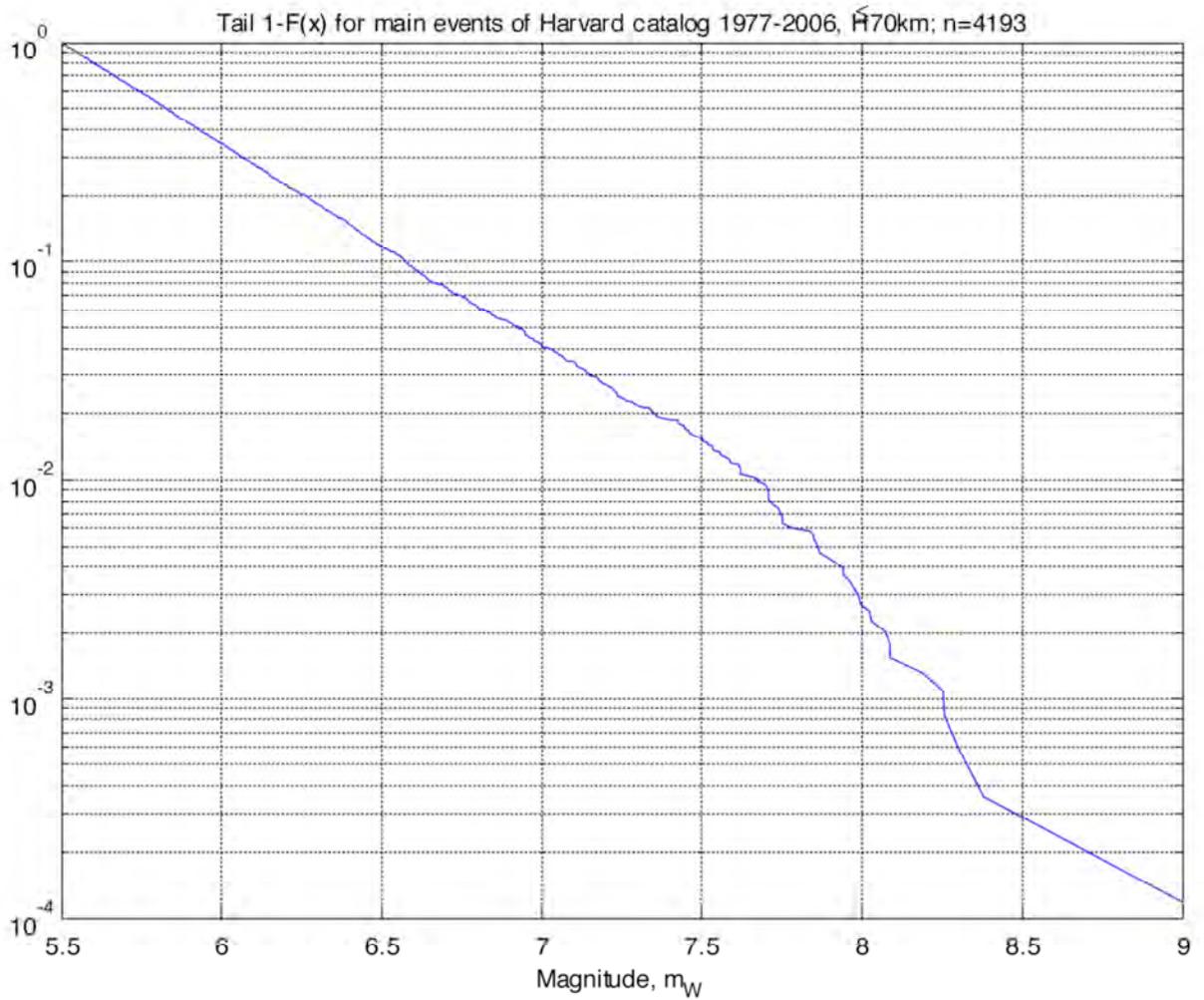

Fig. 1: Complementary cumulative distribution (log-scale) of the main shock magnitudes obtained after declustering of the Harvard catalog (01.01.1977 to 16.06.2006), as described in the text.

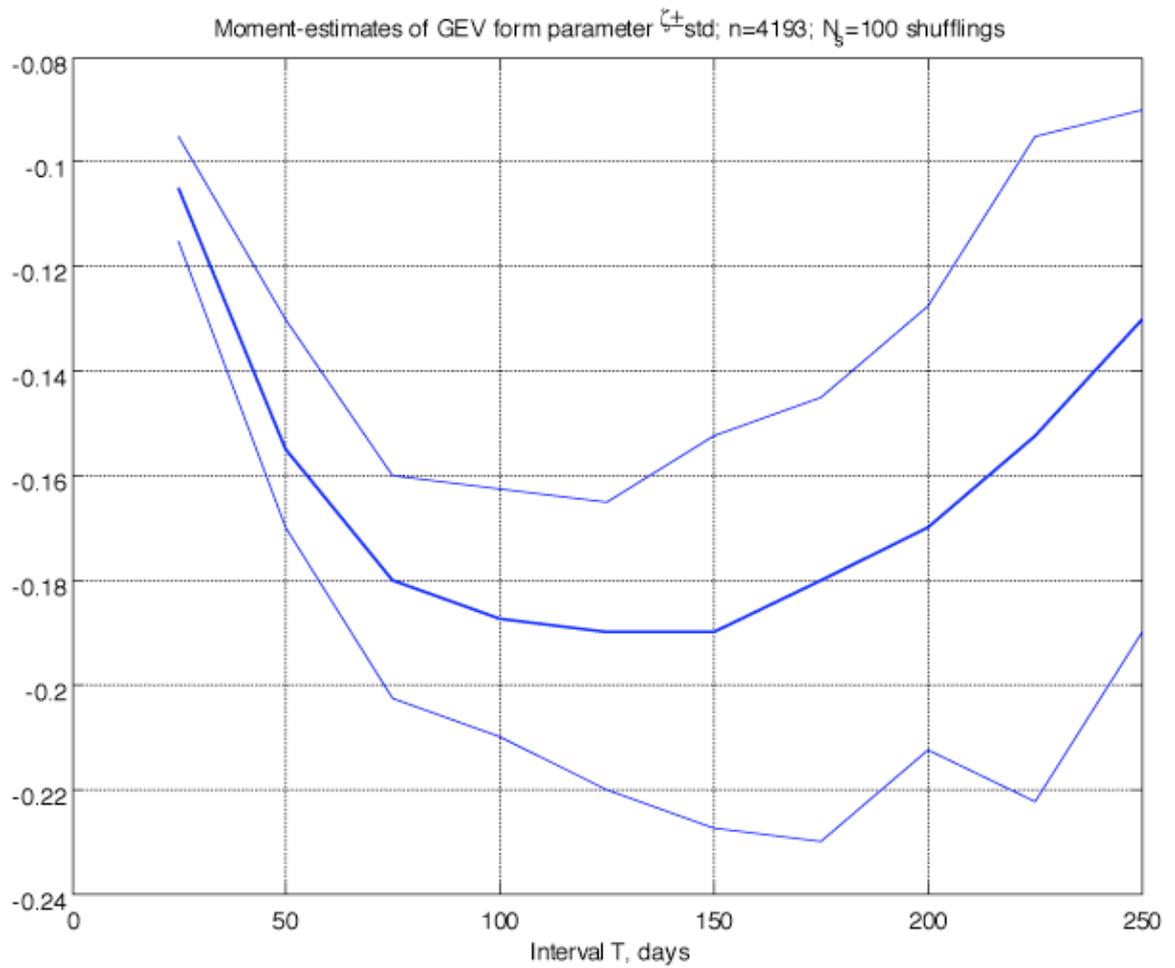

Fig. 2: Dependence as a function of the duration *T* of the time intervals of the Moment estimate of the form parameter ξ of the GEV distribution to the declustered Harvard catalog (01.01.1977 to 16.06.2006), as described in the text. The upper and lower thin lines give the plus-and-minus one standard deviation around the central estimate shown as the thick line. We used 100 shuffling samples, as explained in section 2.5.

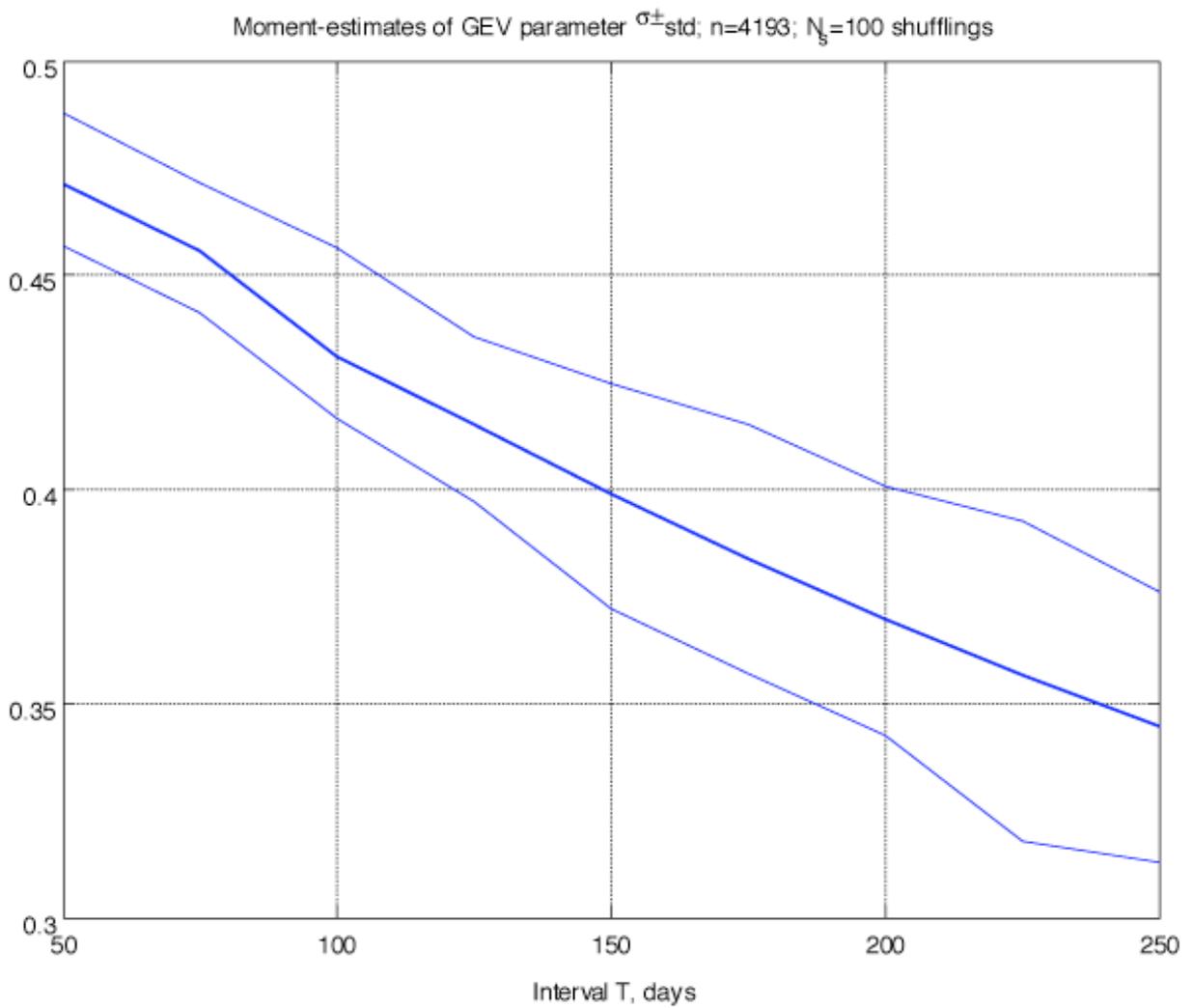

Fig. 3: Dependence as a function of the duration *T* of the time intervals of the Moment estimate of the scale parameter σ of the GEV distribution to the declustered Harvard catalog (01.01.1977 to 16.06.2006), as described in the text. The upper and lower thin lines give the plus-and-minus one standard deviation around the central estimate shown as the thick line. We used 100 shuffling samples, as explained in section 2.5.

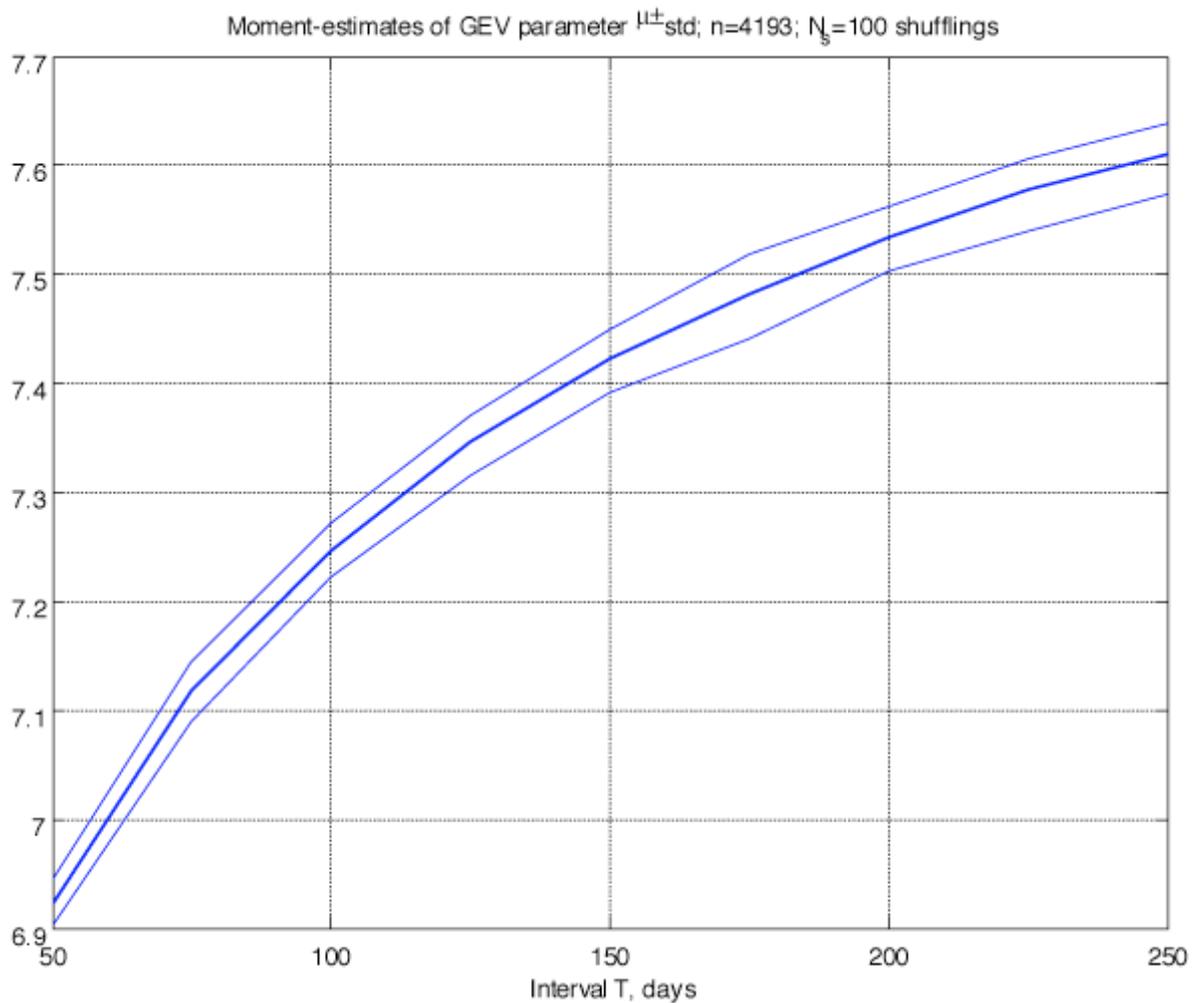

Fig. 4: Dependence as a function of the duration *T* of the time intervals of the Moment estimate of the location parameter μ of the GEV distribution for the declustered Harvard catalog (01.01.1977 to 16.06.2006), as described in the text. The upper and lower thin lines give the plus-and-minus one standard deviation around the central estimate shown as the thick line. We used 100 shuffling samples, as explained in section 2.5.

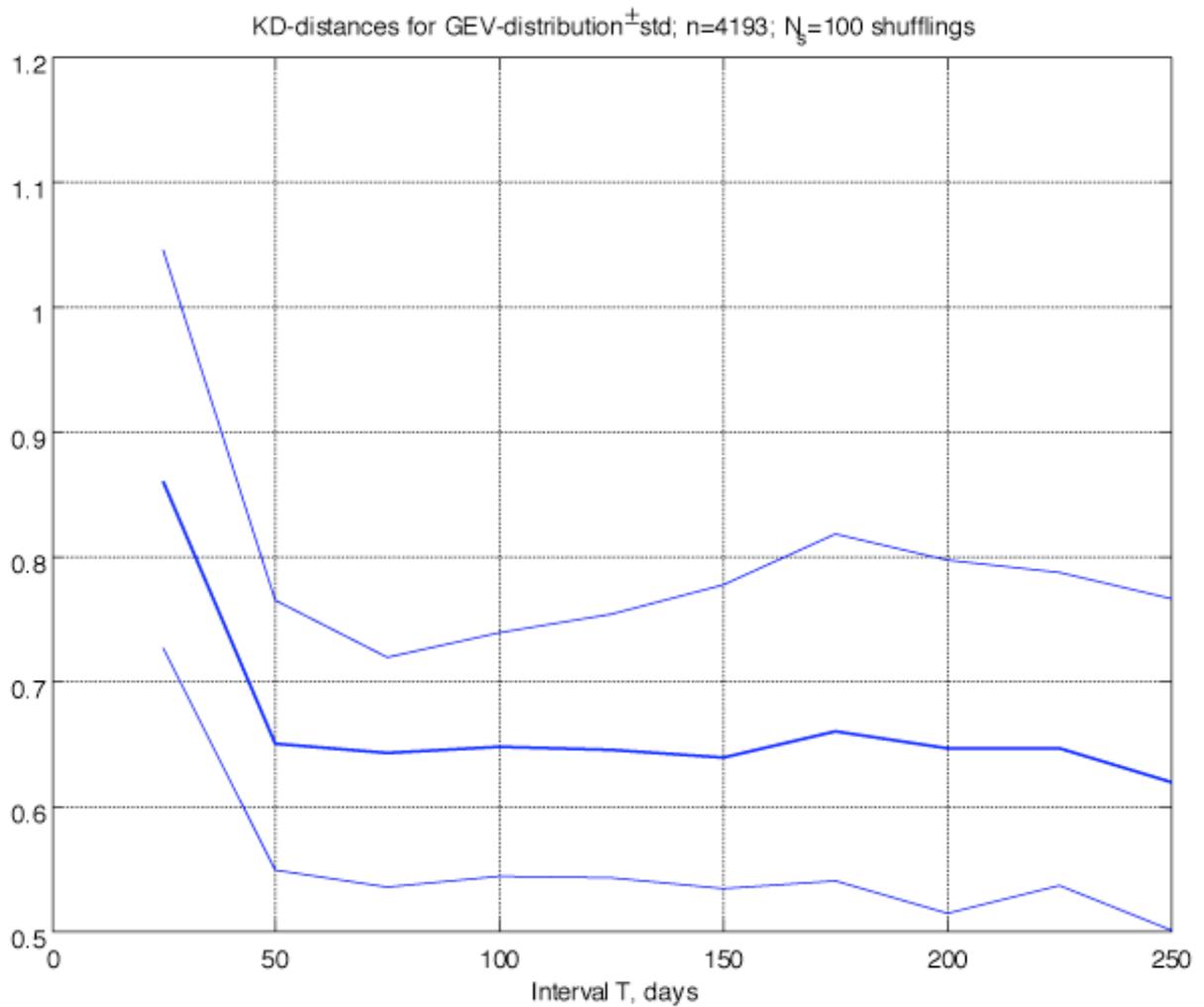

Fig. 5: Kolmogorov distance *KD* (defined in (19)) between the sample distribution of the *T*-maxima and the fitted *GEV*-distribution function as a function of the duration *T* of the time intervals. The upper and lower thin lines give the plus-or-minus one standard deviation of the *KD* value, obtained by simulating *1000* times synthetic *GEV* samples with parameters given by (20) chosen close to our estimates for the declustered Harvard catalog.

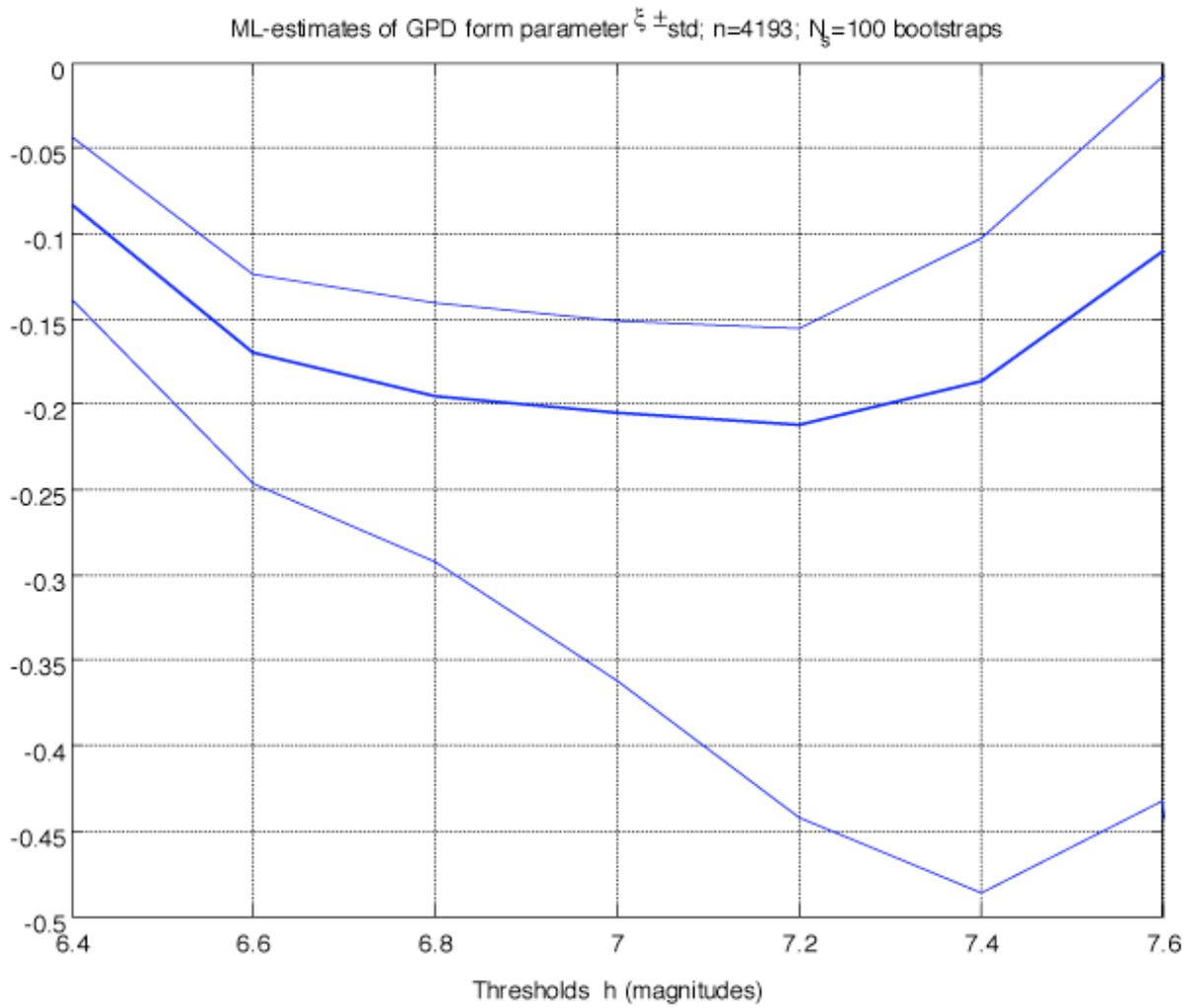

Fig. 6: Dependence as a function of the lower threshold *H* of the estimate of the parameter *ξ*, obtained by the maximum likelihood method with $N_b = 100$ bootstrap samples, for the declustered Harvard catalog.

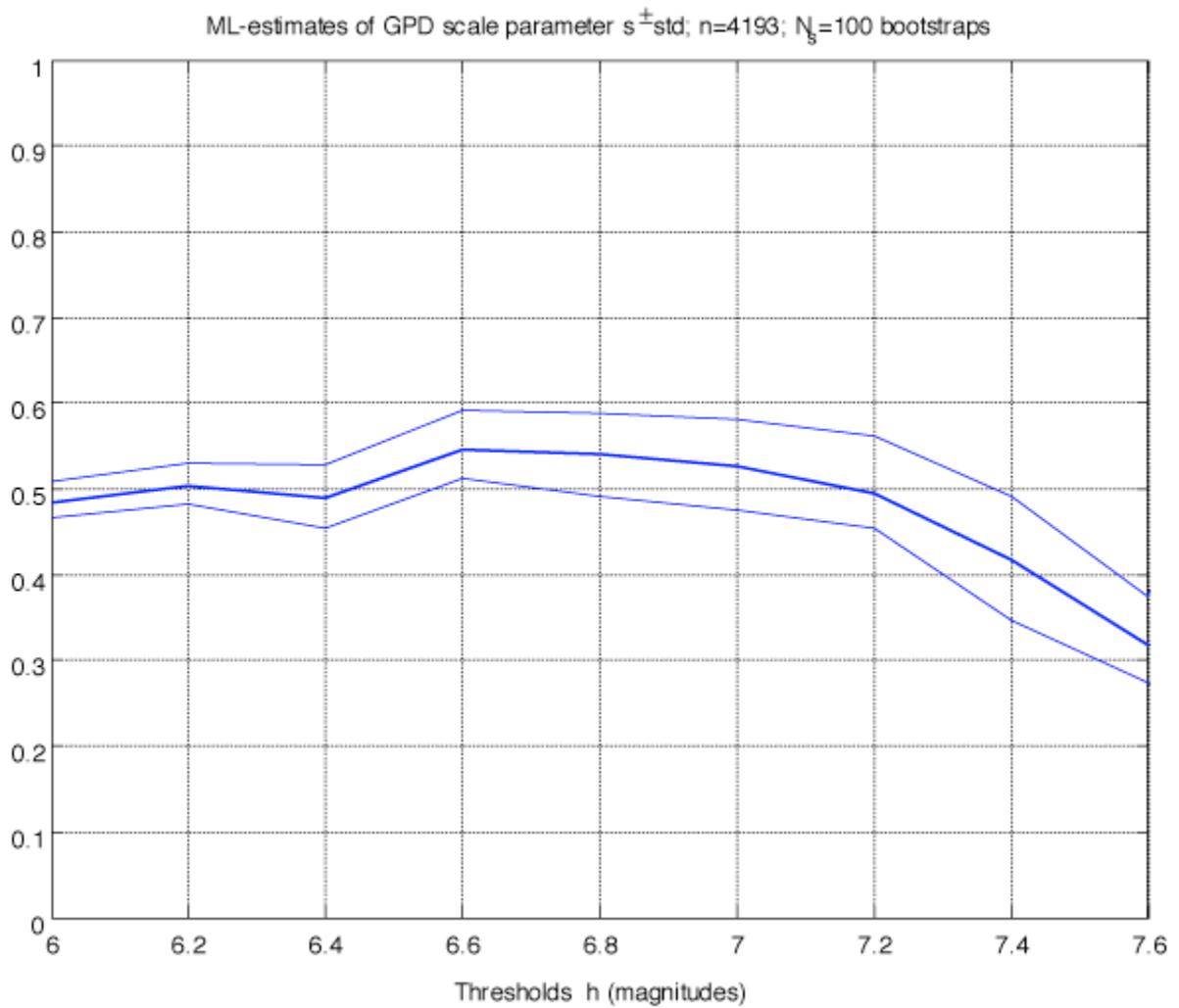

Fig. 7: Dependence as a function of the lower threshold *H* of the estimate of the parameter *s,* obtained by the maximum likelihood method with $N_b = 100$ bootstrap samples, for the declustered Harvard catalog.

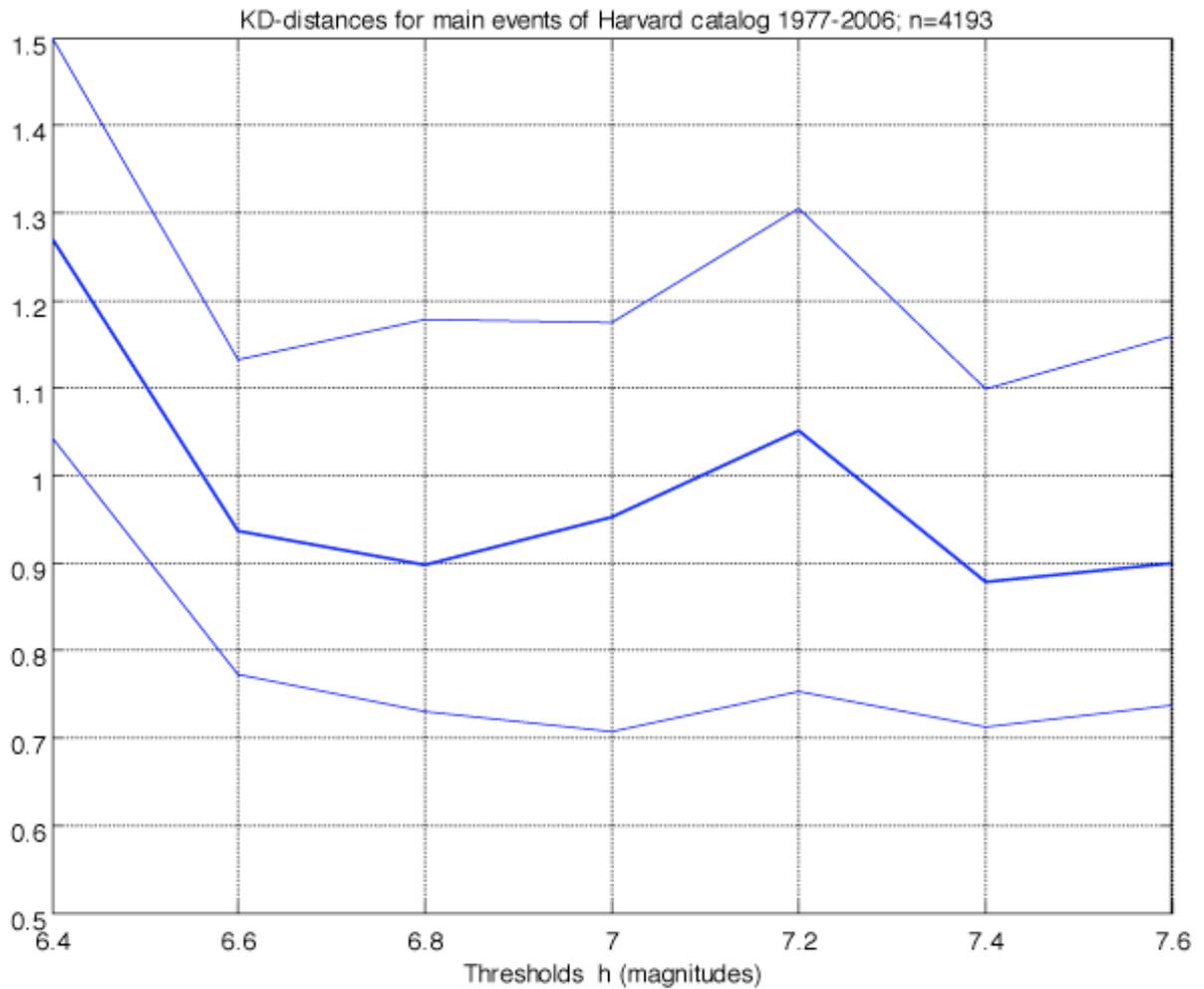

Fig. 8: Dependence as a function of the lower threshold *H* of the median *KD*-distance (thick middle line) defined in equation (26) obtained from *1000* artificial *GPD* samples simulating our real sample taken with the parameters $\xi = -0.2; \, s = 0.53$ close to the values obtained from the empirical data (declustered Harvard catalog).

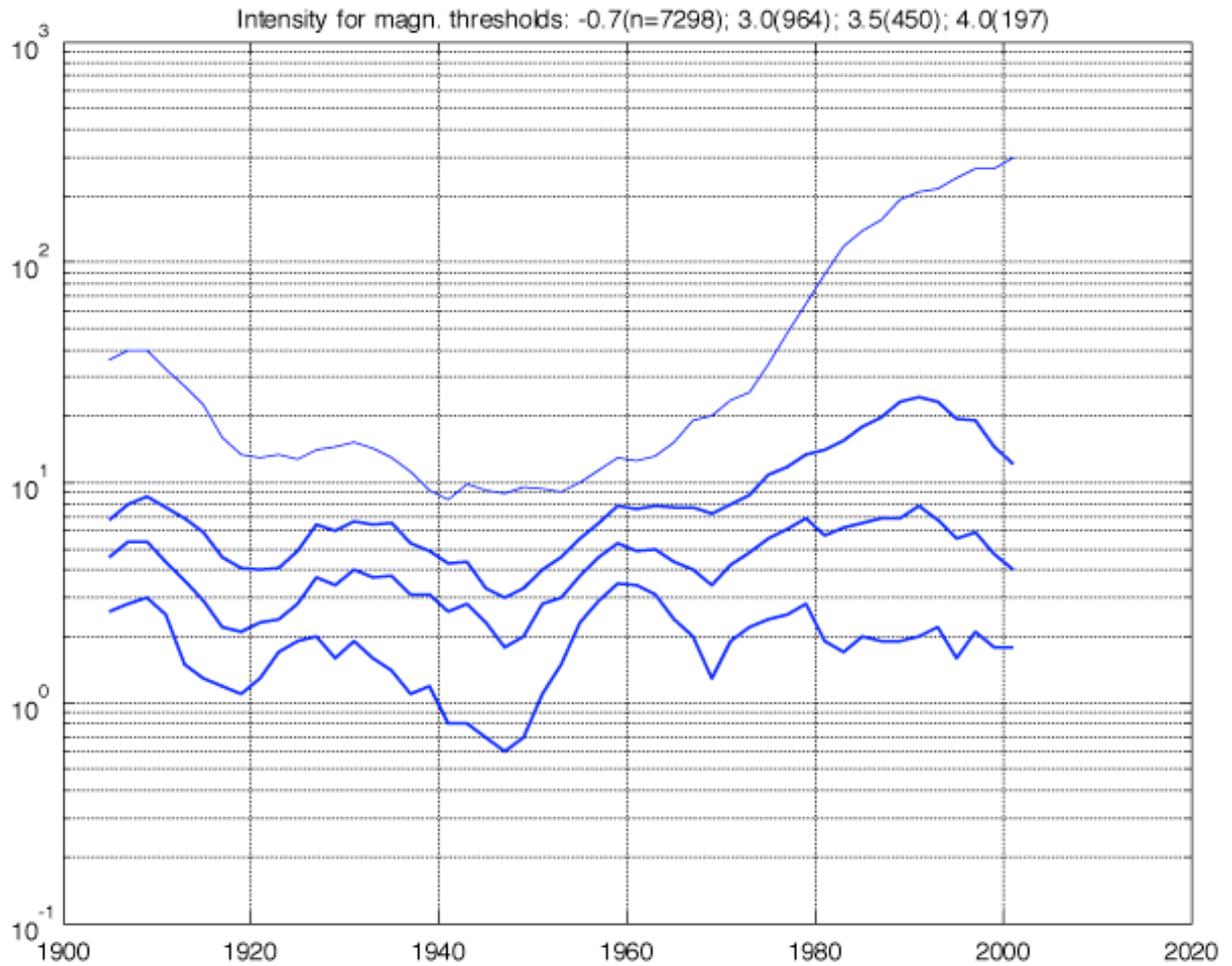

Fig. 9: Intensity of the seismic flow in Fennoscandia over the time period 01.01.1900 – 31.12.2005, defined as the average number of shocks in moving windows of *50* years duration for four lower thresholds (each intensity value is plotted versus the center of the time window).: *m = -0.7 (n = 7298 events, top curve);   m = 3.0  (n = 964 events, second curve from top);   m = 3.5  (n = 450 events, third curve from top);   m = 4.0  (n = 197 events, bottom curve).*

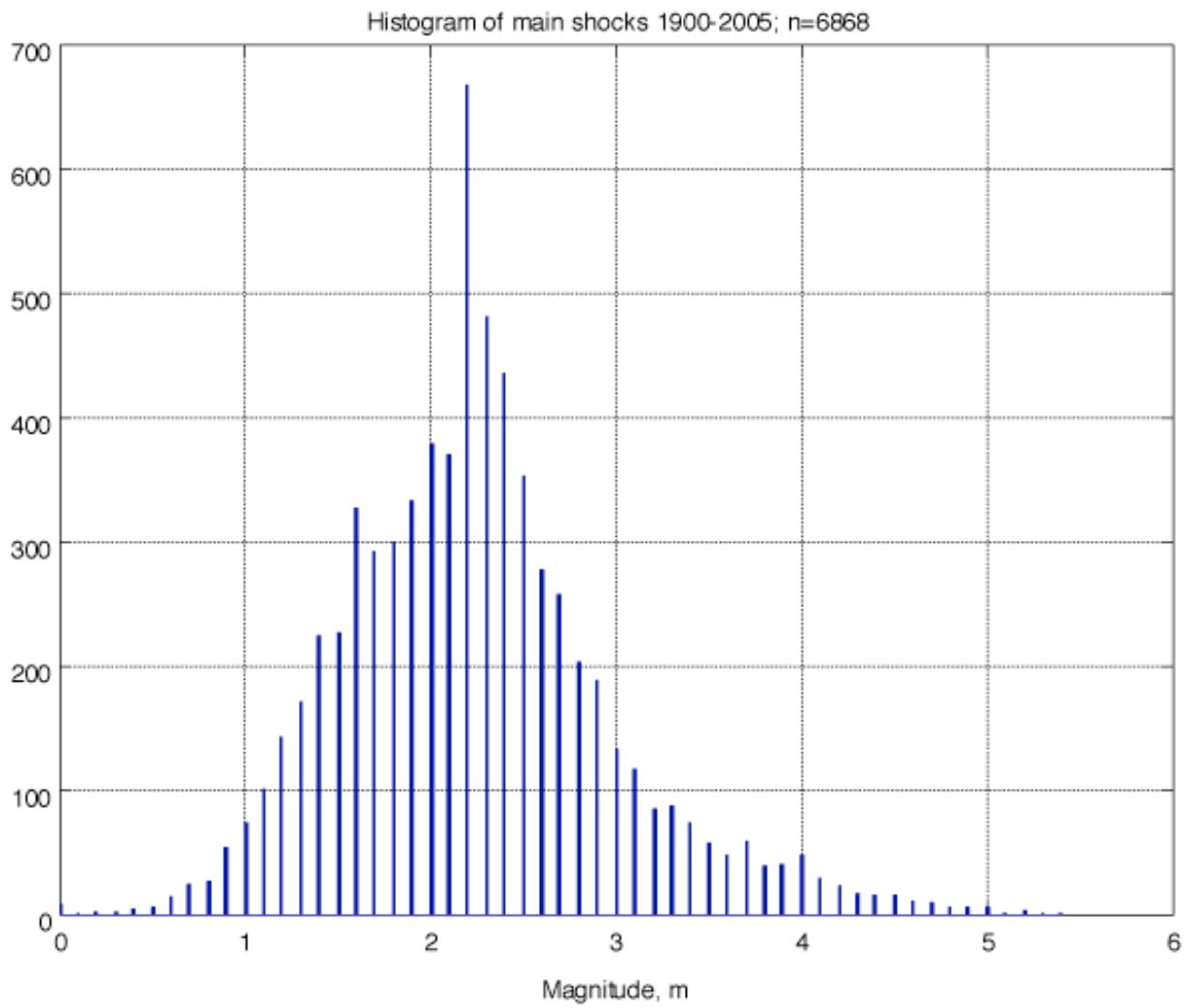

Fig. 10: Histogram of the main shock magnitudes that occurred over Fennoscandia (01.01.1900 – 31.12.2005).

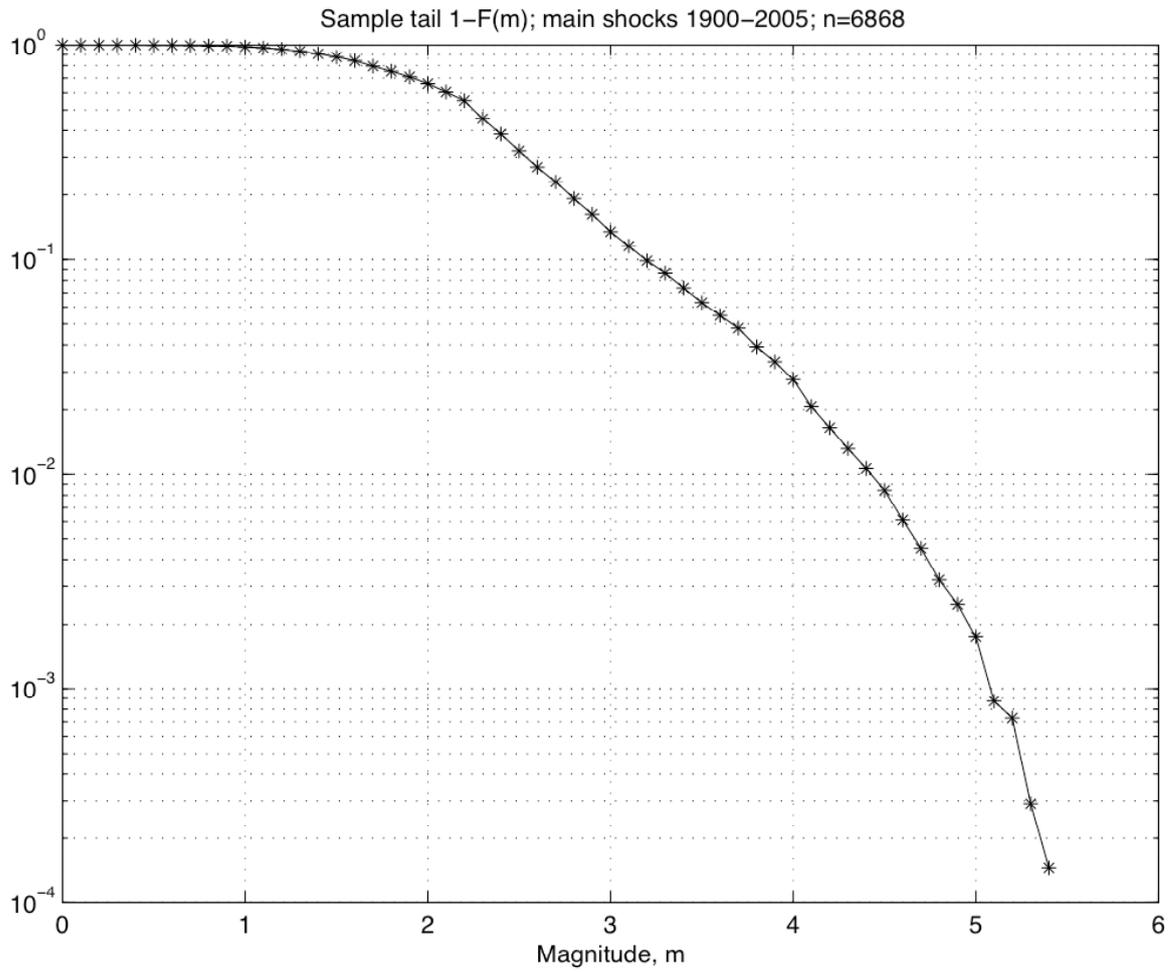

Fig. 11: Sample tail *1 − F(x)* of the main shock magnitudes that occurred over Fennoscandia (01.01.1900 − 31.12.2005).

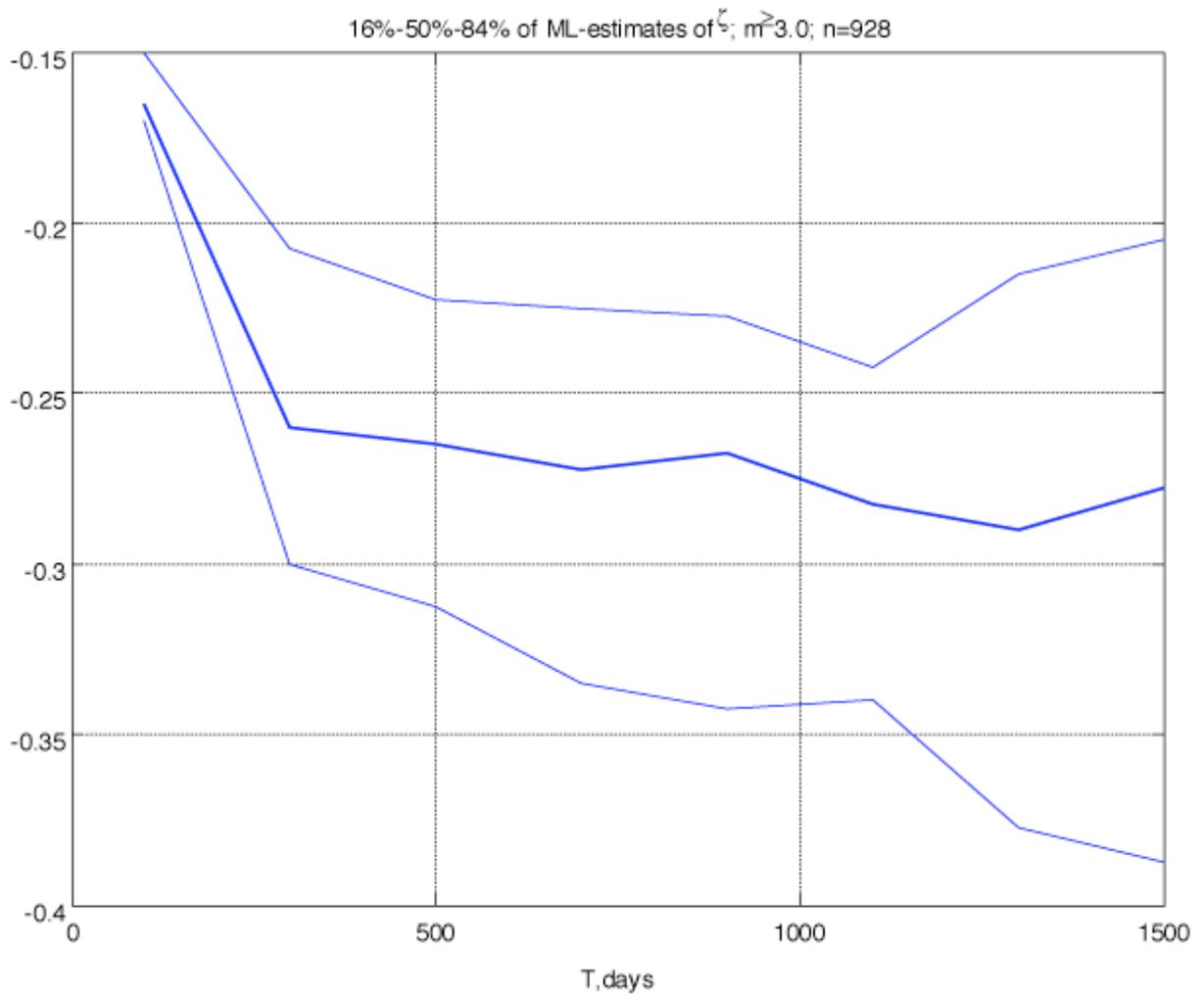

Fig. 12: Moment-estimate of the *GEV*-parameter ξ as a function of *T*, for main shock magnitudes that occurred over Fennoscandia (01.01.1900 – 31.12.2005).

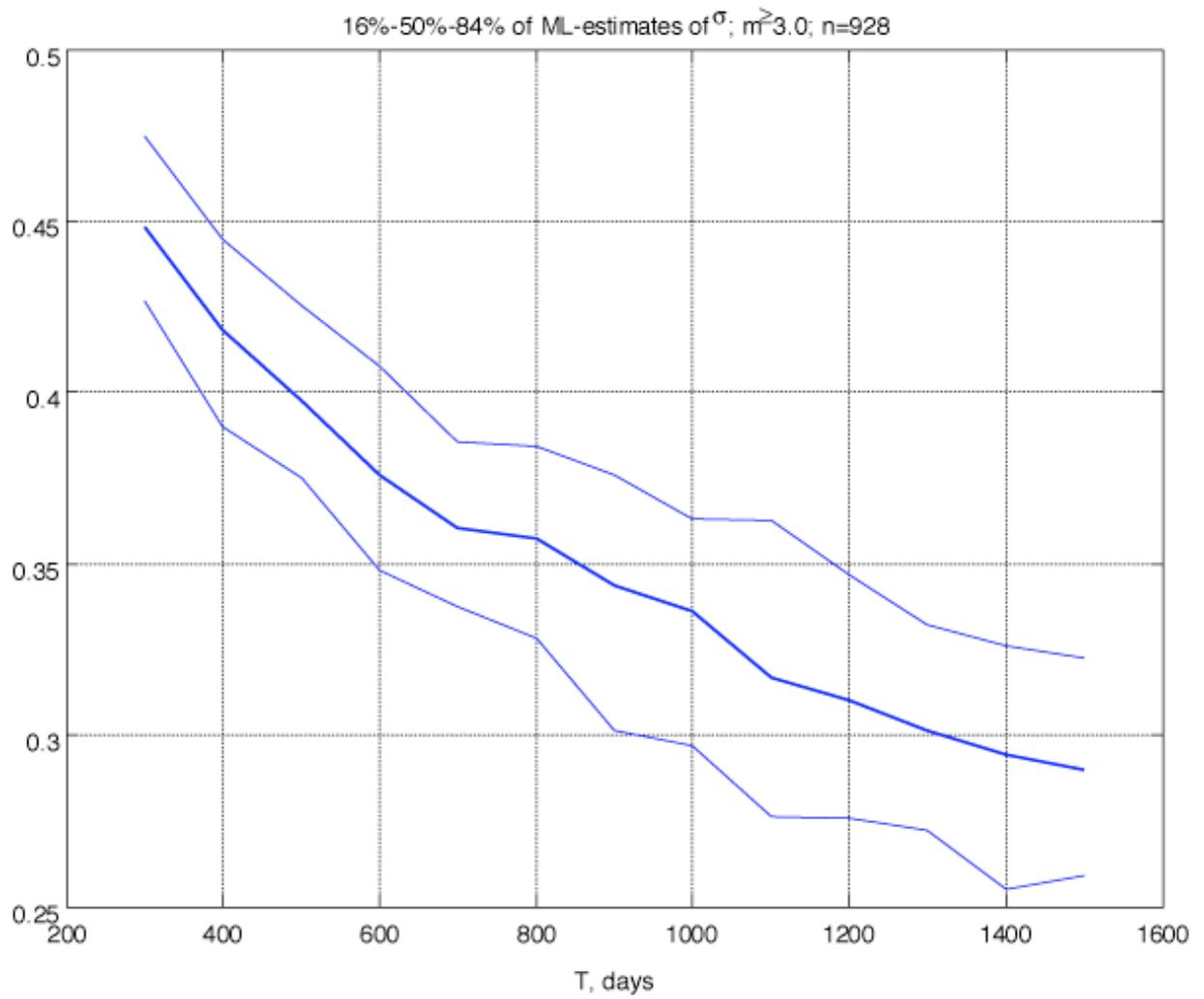

Fig. 13: Moment-estimate of the *GEV*-parameter $\sigma(T)$ as a function of *T*, for main shock magnitudes that occurred over Fennoscandia (01.01.1900 – 31.12.2005).

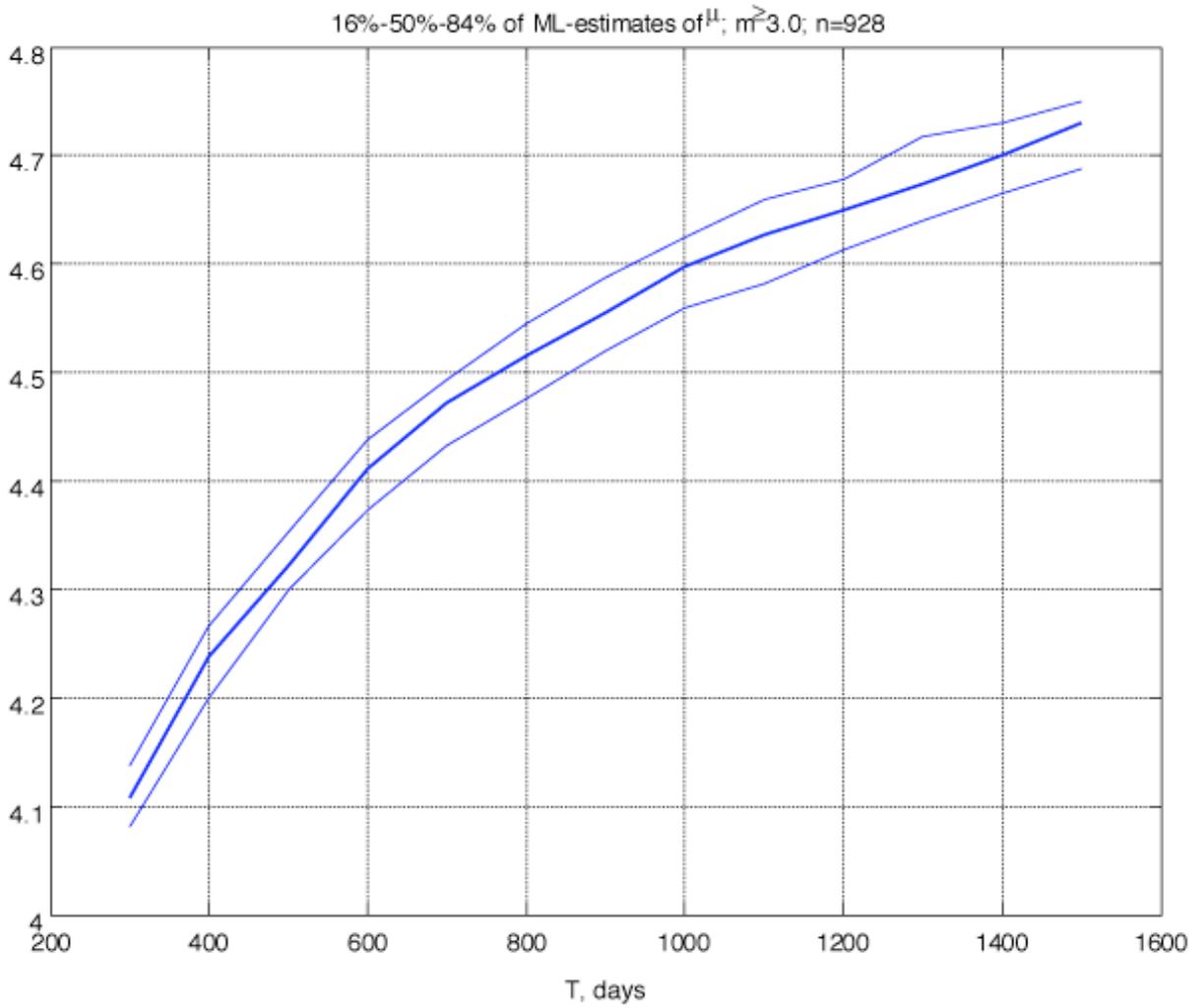

Fig. 14: Moment-estimate of the *GEV*-parameter $\mu(T)$ as a function of *T*, for main shock magnitudes that occurred over Fennoscandia (01.01.1900 – 31.12.2005).

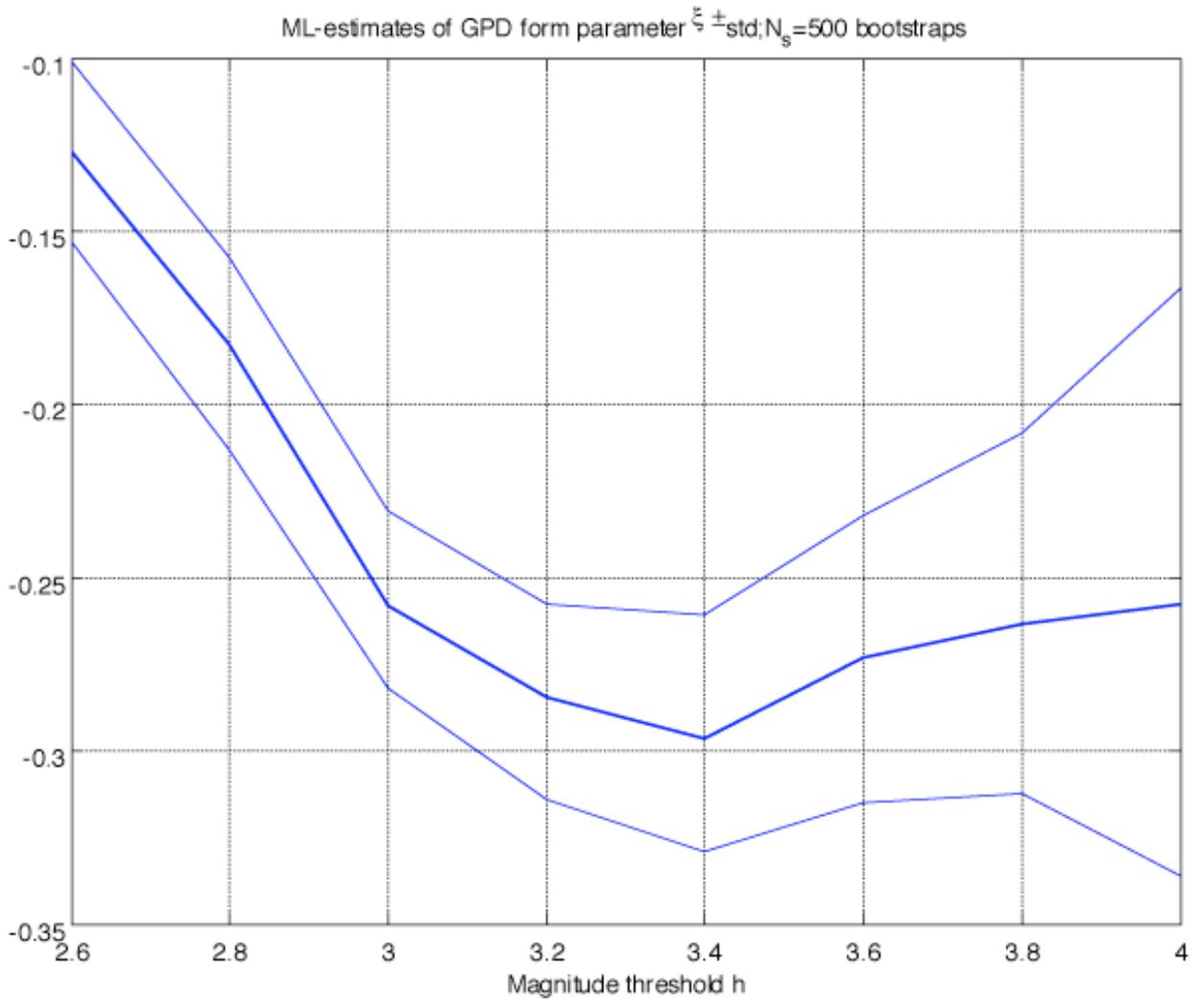

Fig. 15: *ML*-estimate of the form parameter ξ as a function of the lower magnitude threshold *H,* for main shock magnitudes that occurred over Fennoscandia (01.01.1900 – 31.12.2005).

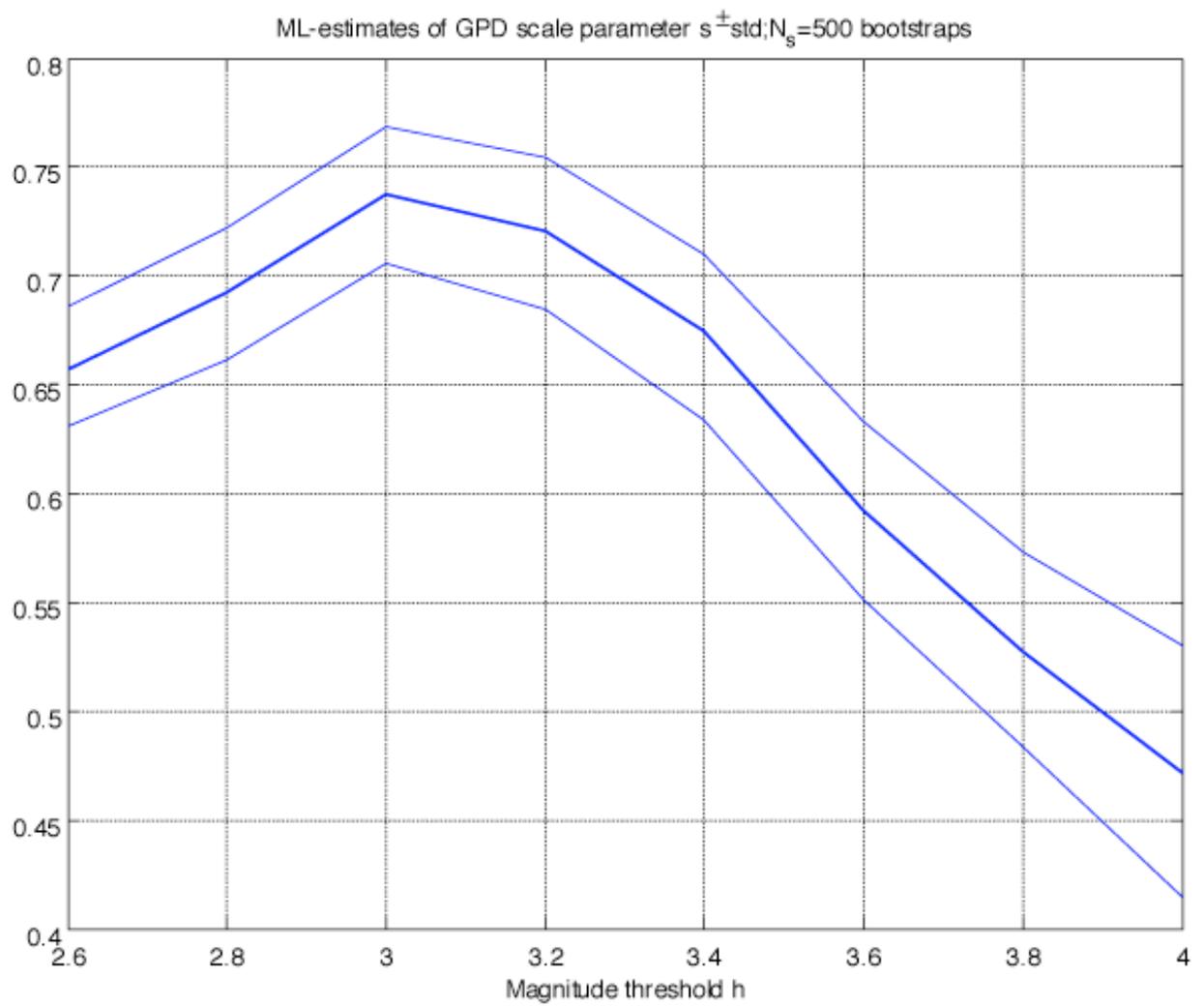

Fig. 16: *ML*-estimate of the scale parameter *s* as a function of the lower magnitude threshold *H*, for main shock magnitudes that occurred over Fennoscandia (01.01.1900 – 31.12.2005).

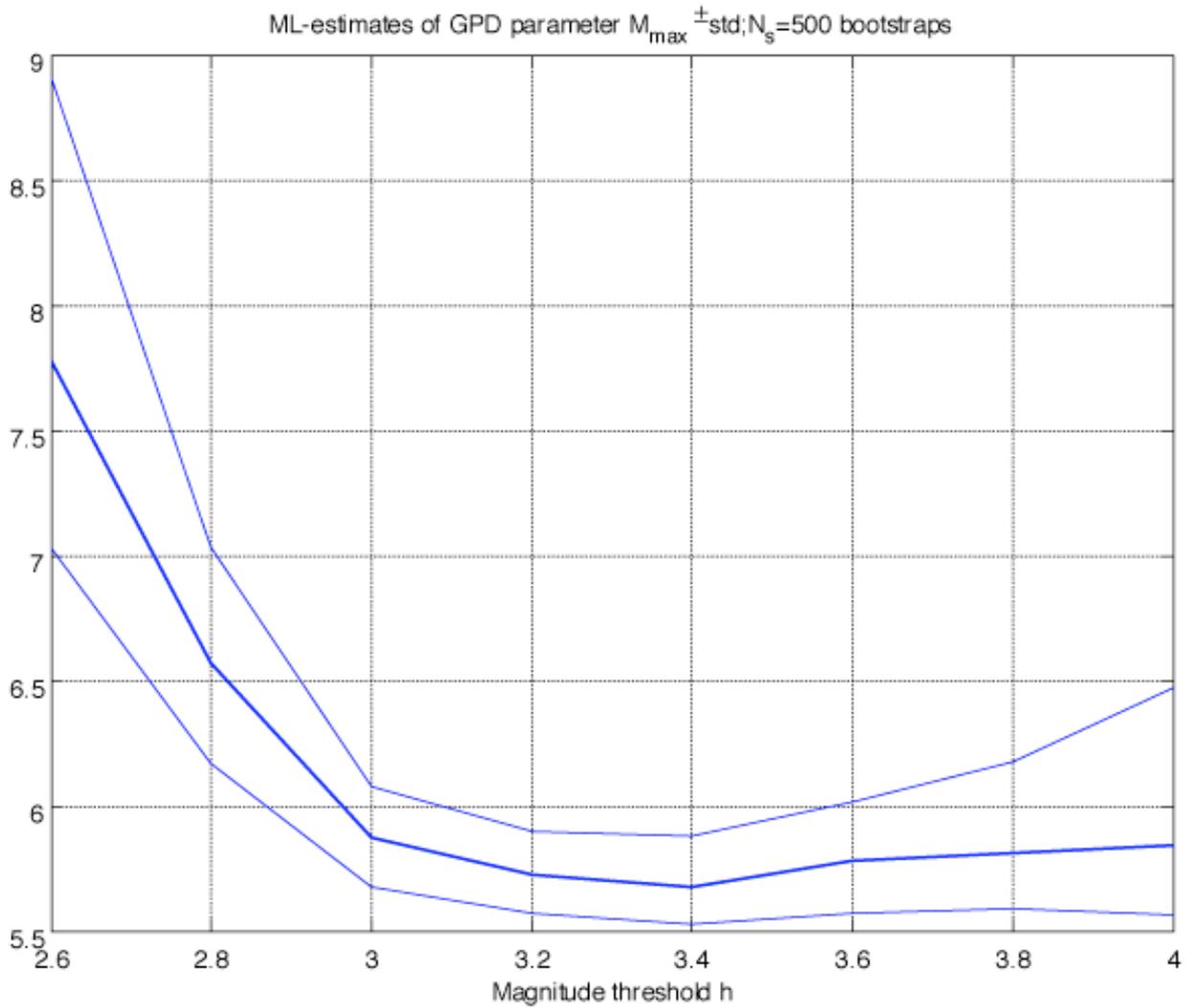

Fig. 17: *ML*-estimate of the tail parameter $M_{max}$ as a function of the lower magnitude threshold *H,* for main shock magnitudes that occurred over Fennoscandia (01.01.1900 – 31.12.2005).